\documentclass[superscriptaddress,oneolumn,secnumarabic,
amssymb,amsmath,nobibnotes,aps,prd,showpacs,nofootinbib]{revtex4}%
\usepackage{graphicx}
\usepackage{epsf}
\usepackage{bm}
\usepackage{amsmath}
\usepackage{amsfonts}
\usepackage{amssymb}
\usepackage{epstopdf}
\usepackage{natbib}
\usepackage{color}%
\providecommand{\U}[1]{\protect\rule{.1in}{.1in}}
\newcommand{\be}{\begin{equation}}
\newcommand{\ee}{\end{equation}}

\newcommand{\mincir}{\raise
-3.truept\hbox{\rlap{\hbox{$\sim$}}\raise4.truept\hbox{$<$}\ }}
\newcommand{\magcir}{\raise
-3.truept\hbox{\rlap{\hbox{$\sim$}}\raise4.truept\hbox{$>$}\ }}

\usepackage{color}
\begin{document}

\title{Novel approach towards the large-scale stable Interacting Dark-Energy models and their Astronomical Bounds}

\author{Weiqiang Yang}
\email{d11102004@163.com}
\affiliation{Department of Physics, Liaoning Normal University, Dalian, 116029, P. R. China}

\author{Supriya Pan}
\email{span@iiserkol.ac.in}
\affiliation{Department of Physical Sciences, Indian Institute of Science Education and Research, Kolkata, Mohanpur$-$741246, West Bengal, India}
\affiliation{Department of Mathematics, Raiganj Surendranath Mahavidyalaya, Sudarshanpur, Raiganj, West Bengal 733134, India}

\author{David F. Mota}
\email{mota@astro.uio.no}
\affiliation{Institute of Theoretical Astrophysics, University of Oslo,
0315 Oslo, Norway}

%%%%%%%%%%%%%%%%%%%%%%%%%%%%%%%%%%%%%%%%%%%%%%%%%%%%%%%%%%%%%%%%%%%%%%%%%%%%%%%%%%%%%%%%%%%%%%%%

\begin{abstract}

Stability analysis of interacting dark energy models generally divides its parameters space into two regions: (i) $w_x \geq -1$ and $\xi \geq 0$ and (ii) $w_x \leq -1$ and $\xi \leq 0$, where $w_x$ is the dark energy equation of state and $\xi$ is the coupling strength of the interaction. Due to this separation, crucial information about the cosmology and phenomenology of these models may be lost.  In a recent study it has been shown that one can unify the two regions with a coupling function which depends on the dark energy equation of state. In this work we introduce a new coupling function which also unifies the two regions of the parameter space and generalises the previous proposal. We analyse this scenario considering the equation of state of DE to be either constant or dynamical. We study the cosmology of such models and constrain both scenarios with the use of latest astronomical data from both background evolution as well as large scale structures. Our analysis shows that a non-zero value of the coupling parameter $\xi$ as well as the dark energy equation of state other than `$-1$' are allowed. However, within $1\sigma$ confidence level, $\xi = 0$, and the dark energy equation of state equal to `$-1$' are compatible with the current data. In other words, the observational data allow a very small but nonzero deviation from the $\Lambda$-cosmology, however, within $1\sigma$ confidence-region the interacting models can mimick the $\Lambda$-cosmology. In fact we observe that the models both at background and perturbative levels are very hard to distinguish form each other and from $\Lambda$-cosmology as well. Finally, we offer a rigorous analysis on the current tension on $H_0$ allowing different regions of the dark energy equation of state which shows that interacting dark energy models reasonably solve the current tension on $H_0$.

\end{abstract}

%%%%%%%%%%%%%%%%%%%%%%%%%%%%%%%%%%%%%%%%%%%%%%%%%%%%%%%%%%%%%%%%%%%%%%%%%%%%%%%%%%%%%%%%%%
\maketitle

\section{Introduction}

Dark matter (DM) and dark energy (DE) are the two heavy constituents of our universe \cite{Ade:2015xua}. The first one, i.e. the dark matter is believed to be responsible for the structure formation of the universe and is almost pressureless while dark energy, a modification of the matter sector in the general relativity, is assumed to be responsible for the current observed acceleration of the universe. Both of them comprise approximately 96\% of the total energy density of the universe with unknown character and origin. Thus, the dynamics of the universe is heavily resting on this sector. Now since the character of this dark sector is unknown, thus, sometimes it is assumed that perhaps the dark matter and dark energy are coupled to each other so that they behave like a single dark fluid. Although this consideration sounds slightly phenomenological but this possibility cannot be ruled out in any ways. From the philosophical point of view we do not have any strong evidence to exclude the phenomenon of dark matter-dark energy interaction. So, one can of course think of some interaction between these two fields. In fact, the standard cosmological laws can be retrieved  at any time under the no interaction limit. Additionally, the dynamics of the universe in presence of any coupling between dark matter and dark energy becomes quite richer with many possibilities. On the other hand, from the theory of particle physics, any two fields can interact with each other. Since both dark energy and dark matter can be thought to be of some fields, for instance some scalar field,
hence, the idea behind the dark matter-dark energy interaction is also supported from the particle physics view point. This henceforth initiated a new branch in the physics of dark energy in the name of interacting dark energy theory.
The idea of coupling in the dark sectors was initiated by Wetterich \cite{Wetterich:1994bg} and subsequently by Amendola \cite{Amendola:1999er}. Thus so far this particular scenario has been explored in the context of current cosmology with some interesting outcomes. The coupling between the dark matter and the dark energy may provide an explanation to the cosmic coincidence problem \cite{Zlatev:1998tr}, a generic problem in the dynamical dark energy models and even in the $\Lambda$-Cold Dark Matter ($\Lambda$CDM) cosmology. Thus, in the last couple of years, a rigorous analysis has been performed by several authors with many interesting possibilities, see for instance \cite{Billyard:2000bh, Olivares:2005tb,delCampo:2008jx,Amendola, Koivisto, delCampo:2008sr,Chimento:2009hj, Quartin:2008px, Valiviita:2009nu, Clemson:2011an, Pan:2013rha, Yang:2014hea, Faraoni:2014vra, Yang:2014gza, Nunes:2014qoa, yang:2014vza,thor,barrow, amendola, llinares, Pan:2014afa, Chen:2011cy, Tamanini:2015iia, Pan:2012ki, Duniya:2015nva, Valiviita:2015dfa, Yang:2016evp, Pan:2016ngu, Mukherjee:2016shl, Sola:2016ecz, Sharov:2017iue, Cai:2017yww, Santos:2017bqm, Mifsud:2017fsy}. The investigations in coupled dark energy have been further fueled  when the latest observational data estimated a nonzero interaction in the dark sector \cite{Salvatelli:2014zta, Nunes:2016dlj,Kumar:2016zpg, vandeBruck:2016hpz, Yang:2017yme, Kumar:2017dnp}. Although most of the studies in this direction automatically assume the interaction between dark matter and dark energy, however, a generalized version of the interacting dark energy is also appealing where the interacting components could be any two barotropic fluids \cite{Barrow:2006hia}.

In this work we propose a novel mechanism to test the stability of interacting dark energy models where the dark energy equation of state can be unrestricted unlike other interacting dark energy models where the state parameter for dark energy is necessarily restricted. To illustrate more in this direction, we consider a spatially flat Friedmann-Lema\^itre-Robertson-Walker universe where pressureless dark matter interacts with dark energy through a nongravitational interaction function. We find that using the pressure perturbation equation for dark energy it is quite possible to construct several interaction models that can be tested without any prior limitation imposed on the dark energy equation of state. The newly constructed interaction functions do have a direct dependence on the dark energy equation of state. However, the entire picture becomes exciting when the interaction function automatically contains such dependence on the dark energy equation of state. We examine the interacting scenario both for constant and dynamical nature of the dark energy equation of state. Finally, we constrain both the interacting scenarios using the latest observational data from different astronomical sources, namely, the cosmic microwave background radiation \cite{ref:Planck2015-1, ref:Planck2015-2}, Baryon Acoustic oscillations distance measurements \cite{Beutler:2011hx,Padmanabhan:2012hf, Manera:2012sc}, Joint Light curve analysis from Supernovae Type Ia \cite{Betoule:2014frx}, weak lensing \cite{Heymans:2013fya,Asgari:2016xuw}, redshift space distortion data \cite{Percival:2004fs,Blake:2011rj,Samushia:2011cs,Reid:2012sw,Beutler:2012px,delaTorre:2013rpa} and finally the Hubble parameter measurements from cosmic chronometers \cite{Moresco:2016mzx} plus the local Hubble constant \cite{Riess:2016jrr}. We note that the use of several observational data provides better and tight constraints on the models.

The paper is organized as follows. In \textit{Section} \ref{sec-2} we describe the background equations of an interacting universe with the new interacting dark energy model with constant and dynamical dark energy equation of state. Thus essentially, we focus on two interacting cosmological scenarions in this study. In \textit{Section} \ref{sec-perturbations} we discuss the interacting cosmology in the perturbative universe. In \textit{Section} \ref{sec-data}, we describe the astronomical data sets that have been used to constrain the models and the \textit{Section} \ref{sec-results} follows the observational constraints of the proposed models. After that in \textit{Section} \ref{sec-comparison} we compare the current models from the statistical ground and also make a comparison with other existing interaction models. The next \textit{Section} \ref{sec-tension} contains an extensive analysis on the current tension on $H_0$. Finally, we close the work in \textit{Section} \ref{conclu} with a short summary of the results obtained.

\section{Interacting cosmology: Background universe}
\label{sec-2}

In this section we describe the governing equations for any
interacting dark energy model. The background geometry is set
to be a spatially flat Friedmann-Lema\^itre-Robertson-Walker (FLRW) universe characterized by
the line element $ds^2 = - dt^2 + a^2 (t) \left[ dr^2 + r^2 \left(d \theta ^2+ \sin^2 \theta d \phi^2\right)\right]$, where $a(t)$ is the expansion scale factor of the universe. The conservation equation of a coupled dark matter and
dark energy system in the FLRW universe can be written as, $\dot{\rho}_c + 3 H \rho_c = - \dot{\rho}_x - 3 H (p_x +\rho_x)$, which can be decoupled into

\begin{eqnarray}
\rho'_c+ 3 \mathcal{H} \rho_c=-aQ,\label{cons1} \\
\rho'_x+ 3 \mathcal{H} (p_x+\rho_x)=aQ,\label{cons2}
\end{eqnarray}
with a new quantity $Q$, known as the interaction rate between the dark sectors. Here $\mathcal{H}=a'/a$ is the conformal Hubble parameter in which the prime denotes the differentiation with respect to the conformal time;
$\rho_c$, $\rho_x$ are respectively the energy densities of pressureless dark matter and dark energy and $p_x$ is the pressure of the dark energy fluid. In addition, we consider the non-relativistic baryons (energy density $\equiv$ $\rho_b$) and relativistic radiation (energy density $\equiv$ $\rho_r$). Since the physics of baryons and radiation are quite well known, so we assume that they are conserved independently, that means the evolution laws of baryons and radiation are $\rho_b \propto a^{-3}$ and $\rho_r \propto a^{-4}$, respectively. The dynamics of the universe is constrained by the Friedmann equation
\begin{eqnarray}
\left(\frac{\mathcal{H}}{a}\right)^2 = \left(\frac{8 \pi G}{3}\right)(\rho_b+\rho_r+ \rho_c+\rho_x),
\end{eqnarray}
which is the constraint equation of the cosmological scenario.  Now, to go ahead one needs a specific functional form for $Q$, and we follow the same tradition. In the literature several forms of $Q$ exist and the most used interactions are $Q \propto \rho_c$, $Q \propto \rho_x$, $Q \propto (\rho_c+\rho_x)$. In this work, we propose a new interaction of the form

\begin{eqnarray}\label{int}
Q=\xi \dot{\rho}_x = \left(\xi/a\right) \rho_x^\prime,
\end{eqnarray}
where the dot represents the derivative with respect to the cosmic time while the prime as mentioned, stands for the derivative with respect to the conformal time and $\xi$ is the coupling parameter. The direction of energy flow that is determined by the sign of $Q$ is dependent on the sign of the coupling parameter $\xi$ as well as with the evolution of $\rho_x$. Precisely, if $\xi> 0$ then $Q > 0$ whenever $\rho_x^\prime > 0$ and on the other hand for $\xi> 0$, one may obtain $Q < 0$ for $\rho_x^\prime < 0$. Similarly for $\xi< 0$, one can also derive the direction of energy flow bwtween the dark sectors from the character of $\rho_x^\prime$. Now, using the conservation equation (\ref{cons2}), the interaction (\ref{int}) can be written in a general form as

\begin{eqnarray}\label{Q}
Q= \frac{3}{a} \left(\frac{\xi}{\xi- 1}\right)\mathcal{H}\, (p_x+ \rho_x)= \frac{3\xi_{e}}{a}\mathcal{H}\, (p_x+ \rho_x),
\end{eqnarray}
where we call $\xi_e = \xi/(1-\xi)$ to be the effective coupling parameter in the interaction scenario. It is clearly seen from eqn (\ref{Q}) that for $\xi = 1$, the interaction $Q$ becomes infinite. However, we remark that $\xi= 1$ represents a very strong interaction in the dark sector which is not allowed by the observational data, see for instance  \cite{Salvatelli:2014zta, Nunes:2016dlj,Kumar:2016zpg, vandeBruck:2016hpz, Yang:2017yme, Kumar:2017dnp}. The interaction (\ref{Q}) has a very special and nice property. For a barotropic equation of state $p_x= w_x \rho_x$, the interaction is linear while on the other hand, for any other complicated equation of state other than the barotropic equation of state, the interaction (\ref{Q}) represents a nonlinear interaction in the dark sector.  Another interesting scenario we observe in the above interaction is as follows. If we the dark energy is assumed to be the cosmological constant, that means $p_x = -\rho_x$, then the coupling becomes zero ($Q= 0$) even if the coupling parameter is non-zero. That means although there exists an interaction, the evolution equations do not change.
In the present work we consider that the dark energy fluid obeys a barotropic equation of state $w_x$ and thus, the interaction (\ref{Q}) can be recast as

\begin{eqnarray}\label{Q-1}
Q = \frac{3}{a} \left(\frac{\xi}{\xi - 1}\right)\mathcal{H}\, (1+w_x) \rho_x\, .
\end{eqnarray}

Now, in order to measure the coupling of the interaction in presence of a dark energy fluid we consider two distinct possibilities: (A) When dark energy equation of state is constant, or (B) The dark energy equation of state is dynamical. For the dynamical dark energy equation of state, we consider the most well known dark energy parametrization known as Chevallier-Polarski-Linder (CPL) parametrization \cite{Chevallier:2000qy, Linder:2002et}, where the equation of state for dark energy is represented by $w_x= w_0 + w_a (1-a)$. Here $w_0$ is the current value of $w_{x}$, i.e. $w_0 = w_{x}$ at $a=1$ (we note that the present value of the scale factor has been set to be unity) and $w_a = dw_x/da$ at $a = 1$.

\section{Interacting cosmology: Perturbed universe}
\label{sec-perturbations}

Now, in order to study the linear perturbations of the interacting dark energy models we introduce the most general scalar mode perturbation, defined by the following metric \cite{ref:Ma1995,ref:Mukhanov1992,ref:Malik2009}

\begin{eqnarray}
ds^2=a^2(\tau) \Bigl[ -(1+2\phi)d\tau^2+2\partial_iBd\tau dx^i+\Bigl((1-2\psi)\delta_{ij}+2\partial_i\partial_jE \Bigr)dx^idx^j \Bigr],
\label{eq:per-metric}
\end{eqnarray}
where $\phi$, $B$, $\psi$ and $E$ are the gauge-dependent scalar perturbations quantities. Let us proceed with the general perturbation equations. We consider the four velocity of any fluid (denoted by `A') as $u_A^{\mu}= a^{-1} (1- \phi, \partial^{i} v_A)$, where $v_A$ is the peculiar velocity potential which in Fourier space is related to the volume expansion $\theta_A = -k^2 (v_A + B)$. Now, since we are considering a coupling between dark matter and dark energy, thus, we have the following constraint $\nabla_{\nu} T_{A}^{\mu \nu} = Q_A$, where $\sum_{A} Q_A = 0$, and $T^{\mu \nu}_A$ is the usual energy-momentum tensor of the fluid $A$. Now, due to the coupling between the dark sectors, the energy flow and the momentum flow take place in general, thus, representing $\tilde{Q}_A$, $F_{A}^{\mu}$ as respectively the energy transfer rate and the momentum transfer rate, relative to the four velocity $u^{\mu}$, following Refs. \cite{ref:Valiviita2008, ref:Majerotto2010, Clemson:2011an}, one can write that

\begin{eqnarray}
Q_{A}^{\mu} = \tilde{Q}_A u^{\mu} + F_A^{\mu},
\end{eqnarray}
where $\tilde{Q}_A = Q_A + \delta Q_A$, $F_A^{\mu} = a^{-1} (0, \partial^{i} f_A)$ in which $Q_A$ is the background interaction  (in fact, $Q_A = Q$) and $f_A$ is the momentum transfer potential. The perturbed interation assumes, $Q_A^0 = -a \left[Q_A (1+\phi)+\delta Q_A \right]$, $Q_A^i= a \partial^i \left[ Q_A (v+B) + f_A\right]$, and finally, the continuity as well as the Euler equations can respectively be written as

\begin{eqnarray}
\delta_A^{\prime} + 3 \mathcal{H} \left(c_{sA}^2 - w_A \right) \delta_A + 9 \mathcal{H}^2 \left(1+w_A \right) \left(c_{sA}^2- c_{aA}^2 \right)\frac{\theta_A}{k^2} + \left(1+w_A \right) \theta_A -3 \left(1+w_A \right) \psi^{\prime} + \left(1+w_A \right) k^2 \left(B- E^{\prime} \right)\nonumber\\ = \frac{a}{\rho_A} \left(\delta Q_A - Q_A \delta _A \right) + \frac{a Q_A}{\rho_A} \left[\phi + 3 \mathcal{H} \left(c_{sA}^2- c_{aA}^2 \right)\frac{\theta_A}{k^2}\right],\\
\theta_A^{\prime} + \mathcal{H} \left(1-3 c_{sA}^2  \right)\theta_A - \frac{c_{sA}^2}{1+w_A} k^2 \delta_A -k^2 \phi = \frac{a}{(1+w_A)\rho_A} \Bigl[ \left(Q_A \theta -k^2 f_A \right) - \left(1+ c_{sA}^2 \right) Q_A \theta_A \Bigr],
\end{eqnarray}
where we have used the notation $\delta_A = \delta \rho_A/\rho_A$ known as the density contrast, and considered $\pi_A = 0$. The symbols $c_{sA}^2$, $c_{aA}^2$,
are respectively denote the adiabatic and physical sound velocity, $\theta = \theta_{\mu}^{\mu}$ is the volume expansion scalar. We note that in order to escape from any kind of instabilities, we need to assume $c_{sA}^2 \geq 0$. Now, for this specific interaction between dark energy and dark matter, the above perturbation equations can be recast as

\begin{eqnarray}
\delta'_x
&=&-(1+w_x)\left(\theta_x+\frac{h'}{2}\right)
-3\mathcal{H}(c^2_{sx}-w_x)\left[\delta_x+3\mathcal{H}(1+w_x)\frac{\theta_x}{k^2}\right] \nonumber \\
&+&3\mathcal{H}\frac{\xi}{\xi-1}(1+w_x)\left[\frac{\theta+h'/2}{3\mathcal{H}}+3\mathcal{H}(c^2_{sx}-w_x)\frac{\theta_x}{k^2}\right], \\
\theta'_x
&=&-\mathcal{H}(1-3c^2_{sx})\theta_x+\frac{c^2_{sx}}{(1+w_x)}k^2\delta_x
+3\mathcal{H}\frac{\xi}{\xi-1}\left[\theta_c-(1+c^2_{sx})\theta_x\right], \\
\delta'_c
&=&-\left(\theta_c+\frac{h'}{2}\right)
+3\mathcal{H}\frac{\xi}{\xi-1}(1+w_x)\frac{\rho_x}{\rho_c}\left(\delta_c-\delta_x-\frac{\theta+h'/2}{3\mathcal{H}}\right), \label{delta_c}\\
\theta'_c
&=&-\mathcal{H}\theta_c,
\label{eq:perturbation}
\end{eqnarray}
where $\delta\mathcal{H}/\mathcal{H}=(\theta+h'/2)/(3\mathcal{H})$, is the perturbed Hubble expansion rate \cite{ref:Gavela2010} in which the prime denotes the derivative with respect to the conformal time.

Now, due to the presence of interaction between dark matter and dark energy, the pressure perturbation equation for dark energy directly includes the interaction rate $Q$, and consequently, the stability of the interaction model becomes sensitive to the specific forms for $Q$. Moreover, the stability of the interaction model is also related to the dark energy equation of state which appears in the pressure perturbation equation for dark energy as \cite{ref:Gavela2009}

\begin{eqnarray}
\delta p_{x}
=c_{sx}^{2}\delta \rho _{x}+3\mathcal{H}\rho
_{x}(1+w_{x})(c_{sx}^{2}-c_{ax}^{2})\left[ 1-\frac{aQ}{3\mathcal{H}\rho
_{x}(1+w_{x})}\right] \frac{\theta _{x}}{k^{2}}~,
\label{eq:deltap}
\end{eqnarray}
from which one can define the doom factor \cite{ref:Gavela2009}: $d\equiv-aQ/[3\mathcal{H}\rho_x(1+w_x)]$  which analyzes the stability of the concerned interaction model by its sign and
the stability is acheived for $d \leq 0$. Now, for any interaction $Q= \xi \bar{Q}$
($\bar{Q} >0$, but $\xi$, the coupling parameter of the interaction
$Q$, is unrestricted in sign), the stability criterion (i.e., $d \leq 0$) provides 
a restriction on the parameters space as either (i) $\xi \geq 0$ and  $(1+w_{x})\geq 0$, or  (ii) $\xi \leq 0$ and $(1+w_{x})\leq 0$. That  means,
in order to test the stability of any interaction, two separate regions must be investigated.  Now, if the interaction contains a factor $(1+w_x)$, that means if $Q = \xi (1+w_x) \bar{Q}$ (where similarly $\bar{Q} > 0$), the doom factor becomes $d \equiv -a\xi \bar{Q}/ ( 3\mathcal{H}\rho_x )$ and it shows that the stability of the interaction is now only characterized by the sign of $\xi$ only. That means the inclusion of the factor $(1+w_x)$ into the interaction releases the prior on $w_x$ which is a beautiful result because now the restriction on the parameter $w_x$ is withdrawn and hence one can test the stability of such interaction models for all $w_x$. This observation has been noticed in a recent paper \cite{ypb}, however, the current work is different. Here, we do not need to include the extra factor $(1+w_x)$ because the interaction that we propose in this work already contains such factor and hence the prior on the dark energy equation of state, $w_x$, automatically released.
To show this, we consider the doom factor for the current interaction (\ref{Q}) which becomes $d = \xi/(1-\xi)$. Since for $d\leq0$, the perturbation evolutions are stable, thus, we conclude that the present interacting dark energy model will be stable if either $\xi\geq1$ or $\xi\leq 0$. However, the possibility $\xi\geq1$ refers to a strong interaction in the dark sector which is not allowed by the present observational data, so we confine our discussions over $\xi\leq0$.

Finally, we close this section with the introduction of growth of matter perturbations. Under the assumption that dark energy does not cluster on sub-Hubble scale \cite{Clemson:2011an}, one can safely neglect the velocity perturbations of $\delta_x = \delta \rho_x/\rho_x$. This assumption is genuine because during the matter dominated era, the acting of dark energy should be subdominant. Therefore, using eqn (\ref{delta_c})
one can derive the following second order differential equation for the density perturbations of pressureless dark matter

\begin{eqnarray}
&&\delta ''_c+\left[1-3\frac{\xi}{\xi-1}(1+w_x)\frac{\rho_x}{\rho_c}\right]\mathcal{H}\delta '_c
=\frac{3}{2}\mathcal{H}^2\Omega_b\delta_b + \nonumber \\
&&\frac{3}{2}\mathcal{H}^2\Omega_c\delta_c \left\{1+ 2\frac{\xi}{\xi-1}(1+w_x)\frac{\rho_{t}}{\rho_c}\frac{\rho_x}{\rho_c}
\left[ \frac{\mathcal{H}'}{\mathcal{H}^2}+1-3w_x+3\frac{\xi}{\xi-1}(1+w_x)\left(1+\frac{\rho_x}{\rho_c}\right)+\frac{w'_x}{\mathcal{H}(1+w_x)} \right] \right\}.
\label{eq:deltacprime2}
\end{eqnarray}
where $\mathcal{H}^2 = (8 \pi G a^2/3)\rho_t$, and
$\rho_t$, is the total energy density of the universe, that means, $\rho_t = \rho_r+\rho_b+\rho_c+\rho_x$. The absence of interaction directs the equation (\ref{eq:deltacprime2}) into $ \delta_m ''+\mathcal{H} \delta_m ' = 4 \pi G \rho_m \delta_m$ ($\rho_m = \rho_c +\rho_b$). Certainly, it is evident that the evolution of $\delta_c$ without any interaction is affected in presence of any interaction as shown in eqn (\ref{eq:deltacprime2}). 
One natural choice is to measure the expansion history due to the interaction, denoted by $\mathcal{H}_{eff}$ and it can be calculated as
\begin{eqnarray}
\frac{\mathcal{H}_{eff}}{\mathcal{H}}=1-\frac{3 \xi}{\xi-1}(1+w_x)\frac{\rho_x}{\rho_c},
\label{eq:Heff}
\end{eqnarray}
which shows that for $\xi= 0$, that when there is no interaction between dark matter and dark energy, $\mathcal{H}_{eff} = \mathcal{H}$ \footnote{We one may note that $\frac{\mathcal{H}_{eff}}{\mathcal{H}}=\frac{ H_{eff}}{H}$. }. Moreover, one can also see that for $w_x = -1$, that means when
cosmological constant interacts with dark matter, $\mathcal{H}_{eff} = \mathcal{H}$.
While on the other hand, there is another quantity, namely the
gravitational constant, $G$, that is also modified in presence of the interaction. We denote the modified gravitational constant as $G_{eff}$, defined by

\begin{eqnarray}
\frac{G_{eff}}{G}=1+ \frac{2 \xi}{\xi-1}(1+w_x)\frac{\rho_{t}}{\rho_c}\frac{\rho_x}{\rho_c}
\left[ \frac{\mathcal{H}'}{\mathcal{H}^2}+1-3w_x+\frac{3 \xi}{\xi-1}(1+w_x)\left(1+\frac{\rho_x}{\rho_c}\right)+\frac{w'_x}{\mathcal{H}(1+w_x)} \right].
\label{eq:Geff}
\end{eqnarray}
It is easy to see that in absence of any coupling, i.e. when $\xi = 0$, $G_{eff} = G$, as expected. Also, for the case of interacting cosmological constant, this equality holds good. From the evolution of $\mathcal{H}_{eff}/\mathcal{H}$, and $G_{eff}/G$, one can quantify the actual deviation of $\mathcal{H}$, $G$, when interaction in the dark sectors is considered. Lastly. we would like to remark on the growth rate of dark matter which is, $f_c \equiv \frac{d}{d\ln a}(\ln \delta_c) $. Since the Euler equation is  modified in presence of the interaction, thus, the dark matter may not follow the geodesics \cite{Koyama:2009gd}. So, this quantity also plays an important role to measure the deviation of the interacting models from the non-interacting one.

\section{Data and the methodology}
\label{sec-data}

In this section we shall shortly describe the astronomical data that have been used to constrain both the interacting dark energy models and the methodology that we use to constrain the interacting scenarios.

\begin{itemize}

\item \textbf{CMB data:} Cosmic microwave background (CMB) data provide tight constraints on the cosmological models. Here we take the CMB data from the latest observations by Planck team
\cite{ref:Planck2015-1, ref:Planck2015-2} that combine the likelihoods $C^{TT}_l$, $C^{EE}_l$, $C^{TE}_l$ in addition to low$-l$ polarization. We shall denote this data by Planck TT,
TE, EE $+$ lowTEB as denoted in \cite{Ade:2015xua}.\newline

\item \textbf{BAO data:} The baryon acoustic oscillation (BAO) data are also powerful to probe the nature of dark energy.  In our analysis we use the estimated ratio $r_s/D_V$ as a `standard ruler' in which $r_s$ is the comoving sound horizon at the baryon drag epoch and $D_V$ is the effective distance determined by the angular diameter distance $D_A$ and Hubble parameter $H$ as $D_V(z)=\left[(1+z)^2D_A(a)^2\frac{z}{H(z)}\right]^{1/3}$. Three different measurements such as, $r_s(z_d)/D_V(z=0.106)=0.336\pm0.015$ from 6-degree Field Galaxy Redshift Survey (6dFGRS) data \cite{Beutler:2011hx}, $r_s(z_d)/D_V(z=0.35)=0.1126\pm0.0022$ from Sloan Digital
Sky Survey Data Release 7 (SDSS DR7) data \cite{Padmanabhan:2012hf}, $r_s(z_d)/D_V(z=0.57)=0.0732\pm0.0012$ from the SDSS DR9 \cite{Manera:2012sc} have been considered.\newline

\item \textbf{JLA sample:} The first data set that proved the existence of some dark energy fluid in the universe sector is the Supernovae Type Ia (SNIa). In the current analysis we use the latest compilation of the SNIa, namely the Joint Light Curves (JLA) sample \cite{Betoule:2014frx} that constain 740 SNIa in the redshift range $z\in[0.01, 1.30]$.\newline

\item \textbf{Weak lensing (WL) data:} We add the weak gravitational lensing
data from blue galaxy sample compliled from  Canada$-$France$-$Hawaii Telescope Lensing Survey
(CFHTLenS) \cite{Heymans:2013fya,Asgari:2016xuw} to other data sets.\newline

\item \textbf{Redshift space distortion (RSD) data:} We use RSD data from different observational surveys
from 2dFGRS \cite{Percival:2004fs}, the WiggleZ \cite{Blake:2011rj}, the SDSS LRG \cite{Samushia:2011cs}, the BOSS CMASS \cite{Reid:2012sw}, the 6dFGRS \cite{Beutler:2012px}, and the VIPERS \cite{delaTorre:2013rpa}. For the measured values of RSD we refer Table I of Ref. \cite{Yang:2014hea}.\newline

\item \textbf{Cosmic Chronometers (CC) data:} We add the Hubble parameter measurements to our analysis. To measure the Hubble 
parameter values at different redshifts, we use the cosmic
chronometers data that have been recently released
in the redshift interval $0 < z < 2$ \cite{Moresco:2016mzx}, and such CC data are 
very powerful to probe the nature of dark
energy due to their model independent character. For a
detailed analysis of the data and the methodology,
we refer to Ref. \cite{Moresco:2016mzx}. \newline

\item \textbf{Local Hubble constant:} Finally, we add the local Hubble constant value
$H_0= 73.02 \pm 1.79$  km/s/Mpc  obtained with 2.4\%
precision by the Riess et al. \cite{Riess:2016jrr}. We denote it
by R16.

\end{itemize}

Now, to constrain the present interacting models we use the likelihood $\mathcal{L}\propto e^{-\chi^2_{tot}/2}$. Here $\chi^2_{tot} = \sum _{i} \chi^2_i$, and $i$ runs over the all data sets that we use, that means CMB (Planck TT, TE, EE $+$ lowTEB), BAO, BAO, WL, RSD, CC and R16. Then we use the cosmoc \cite{Lewis:2002ah}, a markov chain monte carlo simulation to extract the cosmological parameters associted with the model.
Corresponding to each interacting model our parameter space is increased in compared to the $\Lambda$CDM model with minimum dimensional parameter space, see \cite{Barrow:2014opa} for more discussions in this direction. For the interacting model with constant equation of state in dark energy, $w_x$, we have the following eight dimensional parameters space

\begin{align}
\mathcal{P}_1 \equiv\Bigl\{\Omega_bh^2, \Omega_{c}h^2, \Theta_S, \tau, w_x, \xi, n_s, log[10^{10}A_S]\Bigr\},
\label{eq:parameter_space1}
\end{align}
while for the interacting model with dynamical dark energy equation of state $w_x = w_0 + w_a (1-a)$, following nine dimensional parameters space
\begin{align}
\mathcal{P}_2 \equiv\Bigl\{\Omega_bh^2, \Omega_{c}h^2, \Theta_S, \tau, w_0, w_a, \xi, n_s, log[10^{10}A_S]\Bigr\},
\label{eq:parameter_space2}
\end{align}
is considered. We note that in both Eqns. (\ref{eq:parameter_space1}), (\ref{eq:parameter_space2}) the common parameters, $\Omega_bh^2$, $\Omega_{c}h^2$, $\Theta_S$, $\tau$, $n_s$, $A_S$, are respectively identified as the baryons density, cold dark matter density, ratio of sound horizon to the angular diameter distance, optical depth, scalar spectral index, and the amplitude of the initial power spectrum whereas $w_x$, $\xi$ are the model parameters of the parameters space $\mathcal{P}_1$ and $w_0$, $w_a$, $\xi$ are the model parameters of the space $\mathcal{P}_2$.

\section{Results and the interpretations}
\label{sec-results}

We present a detailed analysis of the interacting scenario considering that the equation of state for dark energy, i.e. $w_x$ could be either constant or dynamical. For conveneince we rename the interacting scenario with constant equation of state in DE as IDE 1 while the interacting scenario with dynamical DE assuming the CPL parametrization by IDE 2. In the following subsections we separately discuss both the scenarios.

\subsection{IDE 1: Constant $w_x$}
\label{sec-constant}

\begin{figure}%[tbh]
		\includegraphics[width=11.2cm,height=11cm]{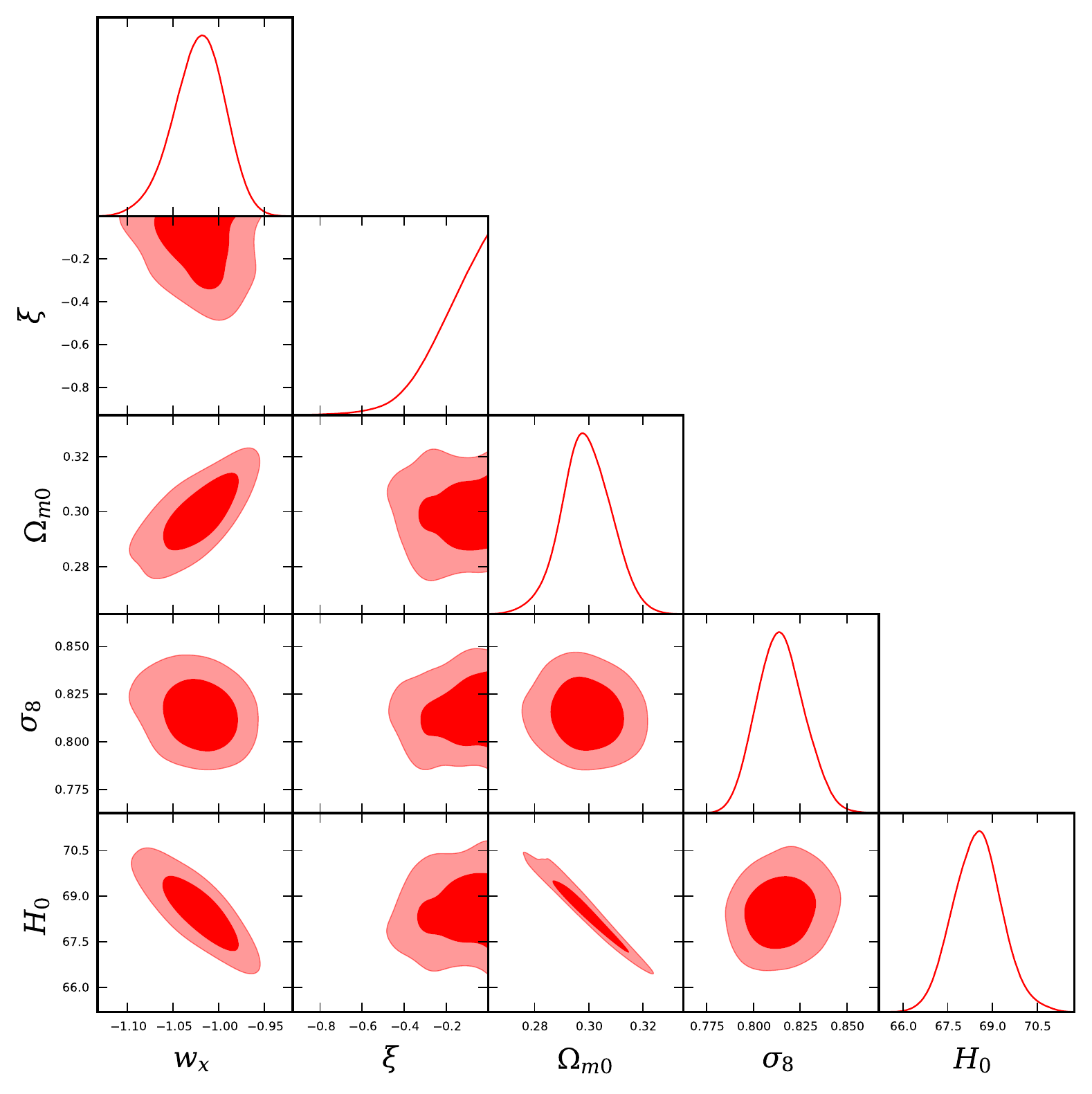}
		\caption{68.3\% and 95.4\% confidence-level contour plots for different combinations of the free parameters of IDE 1 have been shown for the combined analysis Planck TT, TE, EE $+$ lowTEB $+$ BAO $+$ JLA $+$ RSD $+$ WL $+$ CC $+$ R16. }
		\label{fig:contourI}
	\end{figure}

For constant dark energy equation of state, we have analyzed the interacting model with the combined analysis Planck TT, TE, EE $+$ lowTEB $+$ BAO $+$ JLA $+$ RSD $+$ WL $+$ CC $+$ R16. In Table \ref{tab:constantw} we summarize the observational constraints of this scenario for these combined data and the corresponding contour plots at 68.3\% ($1\sigma$), 95.4\% ($2\sigma$) confidence-levels for different combinations of the model parameters along with the 1-dimensional posterior distributions of the model parameters have been displayed in Figure \ref{fig:contourI}.

\begin{figure}%[tbh]
		\includegraphics[width=0.38\textwidth]{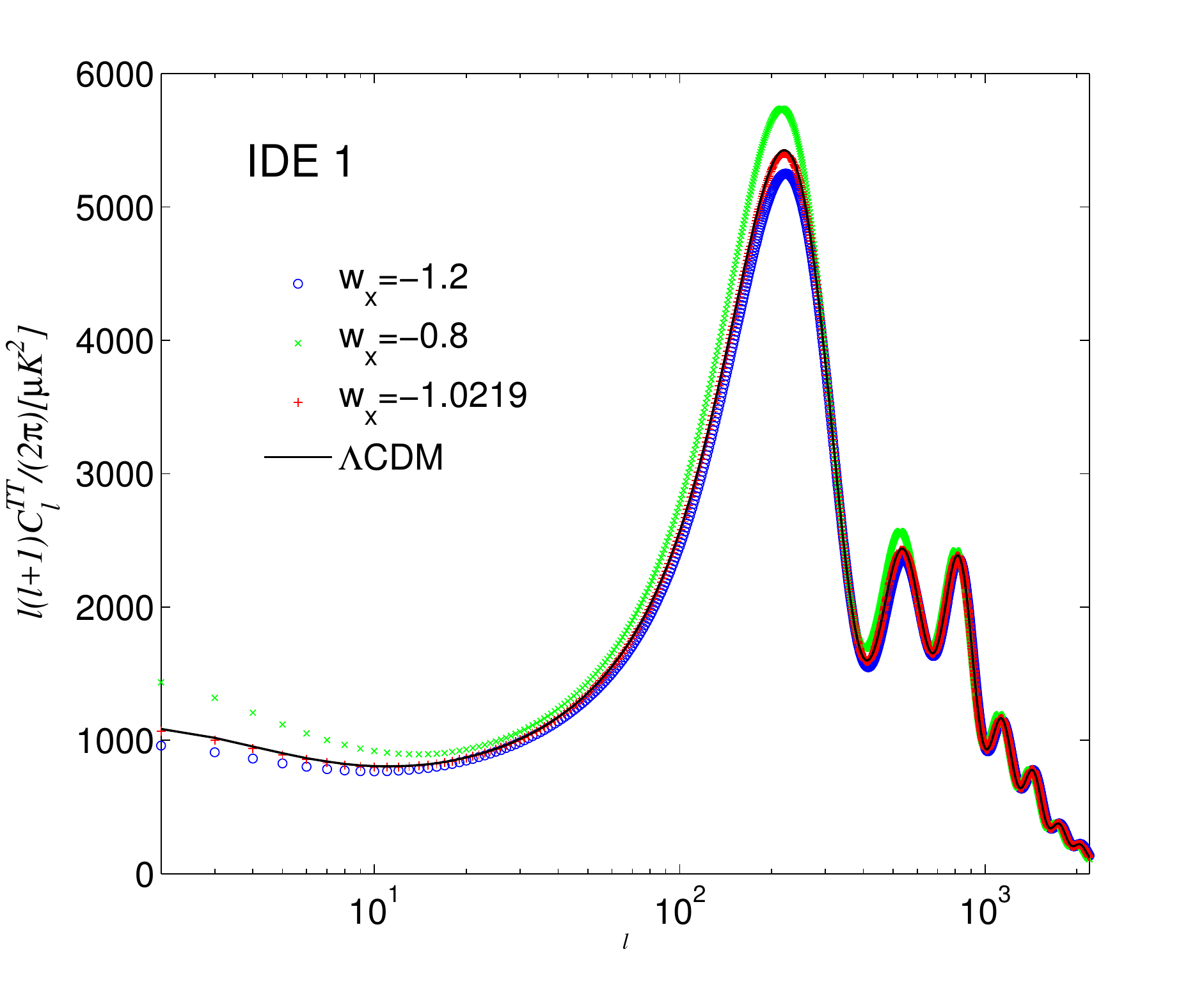}
		\includegraphics[width=0.38\textwidth]{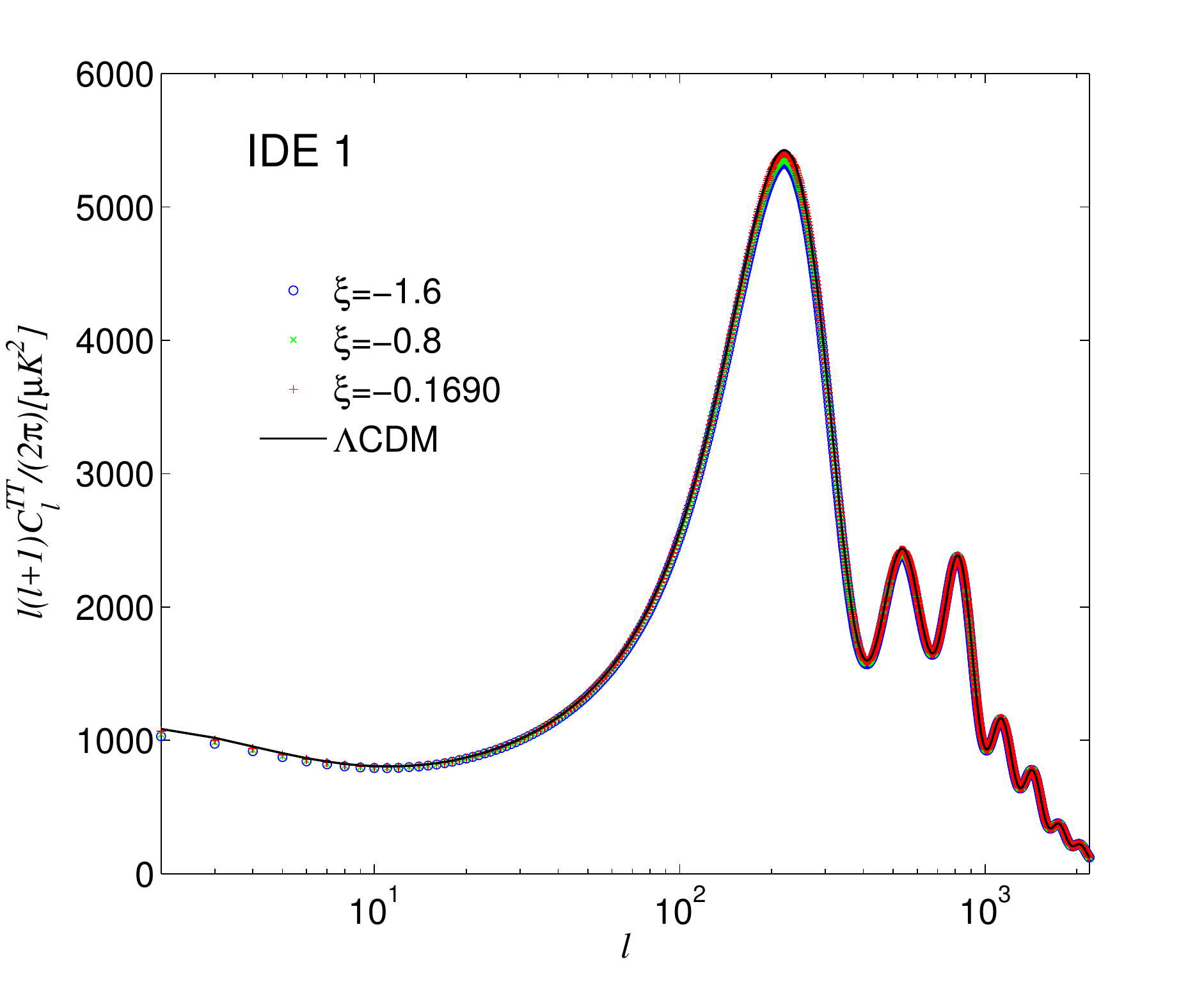}
		\caption{The plots show the angular CMB termperature anisotropy spectra for IDE 1 over the standard $\Lambda$CDM cosmology using the joint analysis data Planck TT, TE, EE $+$ lowTEB $+$ BAO $+$ JLA $+$ RSD $+$ WL $+$ CC $+$ R16. In the left panel we have varied the constant EoS $w_x$ while the right panel stands for diferent values of the coupling parameter $\xi$. We note that the curves in the right panel are so close to each other such that they are practically indistinguishable from each other. }
		\label{fig:cmbplotI}
	\end{figure}

\begin{figure}%[tbh]
		\includegraphics[width=0.33\textwidth]{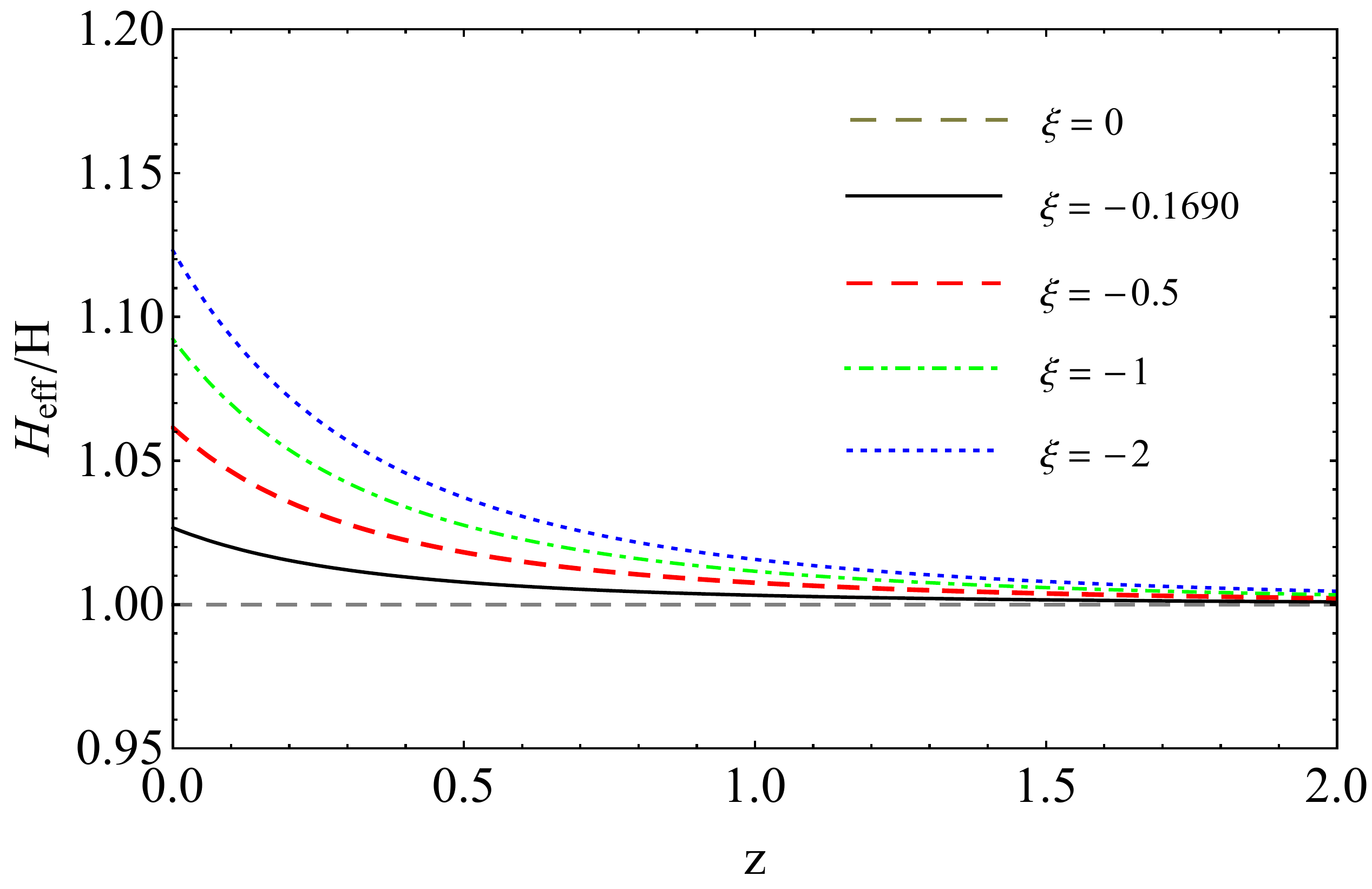}
		\includegraphics[width=0.32\textwidth]{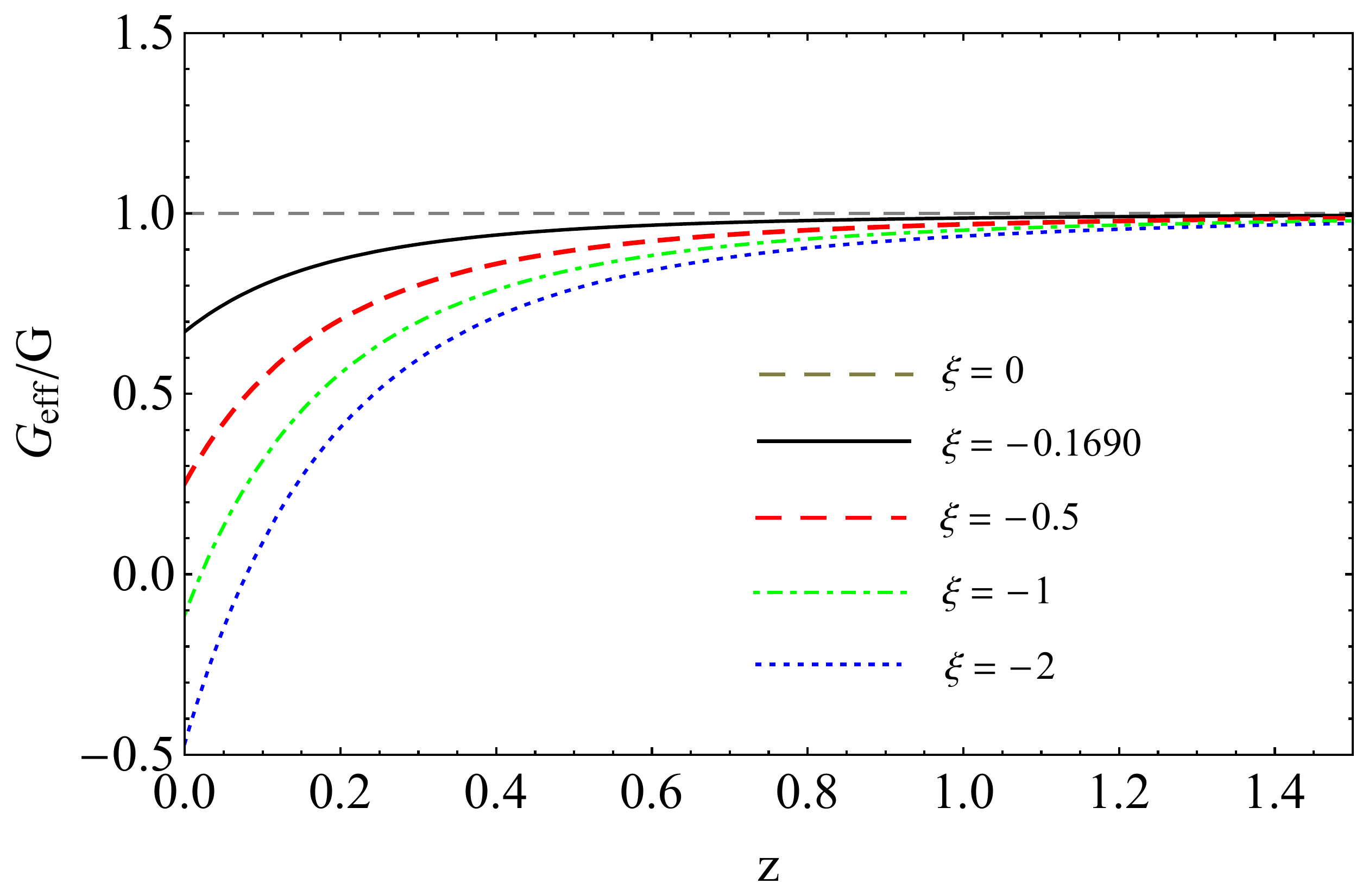}
		\caption{For different coupling parameters, we show the evolutions of $\mathcal{H}_{eff}/\mathcal{H}$ (left panel) and $G_{eff}/G$ (right panel) for the interacting dark energy model with constant $w_x$. The deviation is measured from the non-interacting scenario (i.e. $\xi =0$) and also from the mean value of $\xi = -0.1690$ obtained from the combined analysis Planck TT, TE, EE $+$ lowTEB $+$ BAO $+$ JLA $+$ RSD $+$ WL $+$ CC $+$ R16. }
		\label{fig:new1-cons}
	\end{figure}

From our analysis it is evident that for IDE 1, the coupling parameter as well as the effective coupling parameter are always nonzero. That means the observational data cannot strictly rule out the possibility of a non-interacting scenario. However, within $1\sigma$ confidence level, $\xi = 0$, is compatible according to the present observational data.
Additionally,  it is interesting to note that the best fit value as well as the mode value of the EoS of DE crosses the phantom divide line, however, the analysis also tells that within $1\sigma$ confidence-level, `$w_x = -1$', is consistent with the current astronomical data. Thus, within $1\sigma$ confidence-region, this interaction model can mimick the $\Lambda$-cosmology.  To characterize the large scale behaviour of the IDE models, in Figure \ref{fig:cmbplotI} we display the angular CMB temperature anisotropy spectra in compared to the $\Lambda$CDM cosmology.  In paricular, in the left panel of Fig. \ref{fig:cmbplotI} we show the angular CMB temperature anisotropy spectra for different values of the EoS of DE and in the right panel of Fig. \ref{fig:cmbplotI}, we show the temperature anisotropy of the angular CMB spectra for different values of the coupling parameter. From the left panel of Fig. \ref{fig:cmbplotI} we see that the IDE 1 with $w_x < -1$ does not deviate much from the $\Lambda$CDM cosmology. The deviation is very clear when $w_x > -1$. However, from the right panel of Fig. \ref{fig:cmbplotI}, we see that the small or large coupling basically does not indicate any significant deviation from the $\Lambda$CDM cosmology. This is a surprising result! While from the left panel of Figure \ref{fig:cmbplotI} we see that if $w_x$ increases (i.e. if $w_x$ becomes more quintessential) then we find the deviation in the CMB spectra from that of the standard $\Lambda$CDM cosmology.
Further, in Figure \ref{fig:new1-cons} we depict the evolution of the modified Hubble function $\mathcal{H}_{eff}$ and the modified gravitational constant $G_{eff}$ in presence of the coupling between the dark sectors. The plots in Figure \ref{fig:new1-cons} clearly  show that as the magnitude of the coupling parameter increases, the deviation of both modified Hubble function and the gravitational constant, increases from that of the non-interacting $w_x$CDM model.

\begin{figure}%[tbh]
		\includegraphics[width=0.40\textwidth]{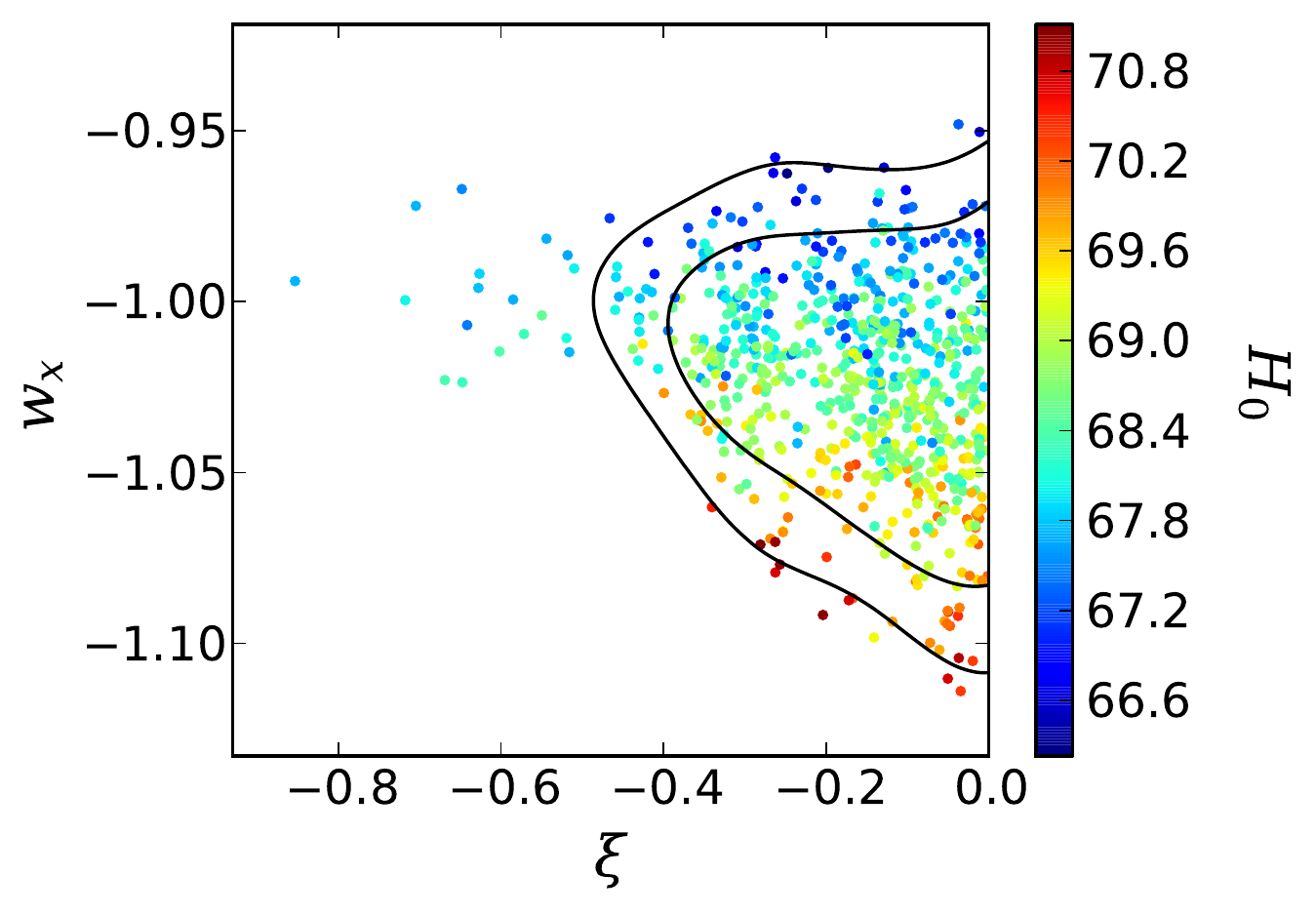}
		\caption{MCMC samples in the $(w_x, \xi)$ plane coloured by the Hubble constant value $H_0$ for IDE 1 analyzed with the combined analysis Planck TT, TE, EE $+$ lowTEB $+$ BAO $+$ JLA $+$ RSD $+$ WL $+$ CC $+$ R16.}
		\label{fig:scatter-IDE1}
	\end{figure}

Finally, we close our analysis with an interesting observation reflected in Figure \ref{fig:scatter-IDE1} which contains the two-dimensional marginalized posterior distribution for the parameters $(w_x, \xi)$. The sample points have been taken from the Markov chain Monte Carlo analysis, and they have been colored by the values of
$H_0$. From this figure we see that the higher values of $H_0$ favor the phantom dark energy while for the lower values of $H_0$ the quintessential dark energy is recommended. Moreover, a shifting of the dark energy equation of state from its quintessential to phantom behavior is observed as $H_0$ descreses. This is one of the interesting observations in this work.

\begingroup
\squeezetable
%\begin{center}
\begin{table}
\caption{The table summarizes the constraints on the free parameters of IDE 1 (IDE with constant $w_x$) for the joint analysis Planck TT, TE, EE $+$ lowTEB $+$ BAO $+$ JLA $+$ RSD $+$ WL $+$ CC $+$ R16. Here, we mention that $\Omega_{m0}= \Omega_{c0}+ \Omega_{b0}$.}
\begin{tabular}{cccccc}
\hline\hline
Parameters & Priors & Mode $\pm$ $1\sigma$ $\pm$ $2\sigma$  & Best fit \\ \hline\hline
$\Omega_c h^2$ & $[0.01, 0.99]$ & $ 0.11773_{-    0.00128-    0.00273}^{+    0.00143+    0.00268}$ & $    0.11725$\\

$\Omega_b h^2$ & $[0.005, 0.1]$ & $0.02228_{-    0.00017-    0.00029}^{+    0.00016+    0.00032}$ &  $    0.02230$\\

$100\theta_{MC}$ & $[0.5, 10]$  & $1.04068_{-    0.00031-    0.00064}^{+    0.00033+    0.00062}$ &   $1.04051$\\

$\tau$ &   $[0.01, 0.8]$ & $0.06593_{-    0.01618-    0.03060}^{+    0.01599+    0.03192}$ &  $    0.07098$\\

$n_s$ & $[0.5, 1.5]$ & $0.97692_{-    0.00404-    0.00777}^{+    0.00389+    0.00787}$ &  $ 0.98041$\\

${\rm{ln}}(10^{10} A_s)$ & $[2.4, 4]$ & $3.06864_{-    0.03129-    0.05905}^{+    0.03149+    0.06361}$ &  $    3.07831$\\
\hline
$w_x$ & $[-2, 0]$  & $-1.01014_{-    0.02583-    0.05786}^{+    0.03101+    0.05702}$ &  $   -1.02844$\\

$\xi$ & $[-1, 0]$ & $-0.00857_{-    0.04001-    0.23056}^{+    0.00857+    0.00857}$ &  $   -0.02682$\\
\hline
$\Omega_{m0}$ & $-$ & $0.29674_{-    0.00877-    0.01902}^{+    0.00959+    0.01877}$ &  $    0.29510$\\

$\sigma_8$ &  $-$ & $0.81393_{-    0.01411-    0.02289}^{+    0.01165+    0.02510}$ &  $    0.819897$\\

$H_0$ &  - & $ 68.60277_{-    0.82512-    1.55987}^{+    0.79166+    1.63707}$ & $   68.82080$\\
\hline
$\chi^2_{min}$ & $-$ & $-$  & $13720.258$\\
\hline
\end{tabular}
\label{tab:constantw}
\end{table}
%\end{center}
\endgroup

%\begin{center}
%\end{center}

%\begin{center}
%\end{center}

%\begin{center}
%\end{center}

%\begin{center}
%\end{center}

\begingroup
\squeezetable
%\begin{center}
\begin{table}
\caption{The table summarizes the constraints on the free parameters of IDE 2 (IDE with dynamical $w_x$) using the joint analysis Planck TT, TE, EE $+$ lowTEB $+$ BAO $+$ JLA $+$ RSD $+$ WL $+$ CC $+$ R16. Here we note that $\Omega_{m0}= \Omega_{c0}+ \Omega_{b0}$.}
\begin{tabular}{cccccccc}
\hline\hline
Parameters & Priors & Mode $\pm$ $1\sigma$ $\pm$ $2\sigma$ & Best fit \\ \hline\hline
$\Omega_c h^2$ &  $[0.01, 0.99]$ & $0.11766_{-    0.00119-    0.00316}^{+    0.00158+    0.00301}$ &  $    0.11410$\\

$\Omega_b h^2$ & $[0.005, 0.1]$ & $0.02224_{-    0.00015-    0.00029}^{+    0.00017+    0.00030}$ &  $    0.02241$\\

$100\theta_{MC}$ &  $[0.5, 10]$ & $ 1.04061_{-    0.00030-    0.00063}^{+    0.00031+    0.00064}$ &  $    1.04088$\\

$\tau$ & $[0.01, 0.8]$ & $0.06443_{-    0.01599-    0.03523}^{+    0.01633+    0.03277}$  &  $    0.06886$\\

$n_s$ &  $[0.5, 1.5]$ & $0.97561_{-    0.00408-    0.00799}^{+    0.00410+    0.00783}$ & $    0.98110$\\

${\rm{ln}}(10^{10} A_s)$ &  $[2.4, 4]$ & $3.06729_{-    0.03077-    0.06602}^{+    0.03133+    0.06551}$  & $    3.07243$\\
\hline
$w_0$ & $[-2, 0]$ &  $-1.04169_{-    0.06889-    0.10442}^{+    0.05331+    0.10939}$ &  $   -1.04824$\\

$w_a$ & $[-3, 3]$ & $0.04941_{-    0.13556-    0.34331}^{+    0.19888+    0.30253}$ &  $    0.11464$\\

$\xi$ & $[-1, 0]$ & $-0.00989_{-    0.03969-    0.34569}^{+    0.00989+    0.00989}$ &  $   -0.45436$\\
\hline
$\Omega_{m0}$ & - & $ 0.29986_{-    0.00888-    0.01795}^{+    0.00987+    0.01713}$  & $    0.29201$\\

$\sigma_8$ & - & $0.82023_{-    0.01489-    0.02599}^{+    0.01291+    0.02687}$ &  $    0.79875$\\

$H_0$ & - & $68.42602_{-    0.87900-    1.51089}^{+    0.74521+    1.53486}$ & $   68.75732$\\
\hline
$\chi^2_{min}$ & $-$ & $-$ &  $13721.564$ \\
\hline
\end{tabular}
\label{tab:dynamicalw}
\end{table}
%\end{center}
\endgroup

\subsection{IDE 2: Dynamical $w_x$}
\label{sec-dynamical}

\begin{figure}%[tbh]
		\includegraphics[width=11.2cm,height=11cm]{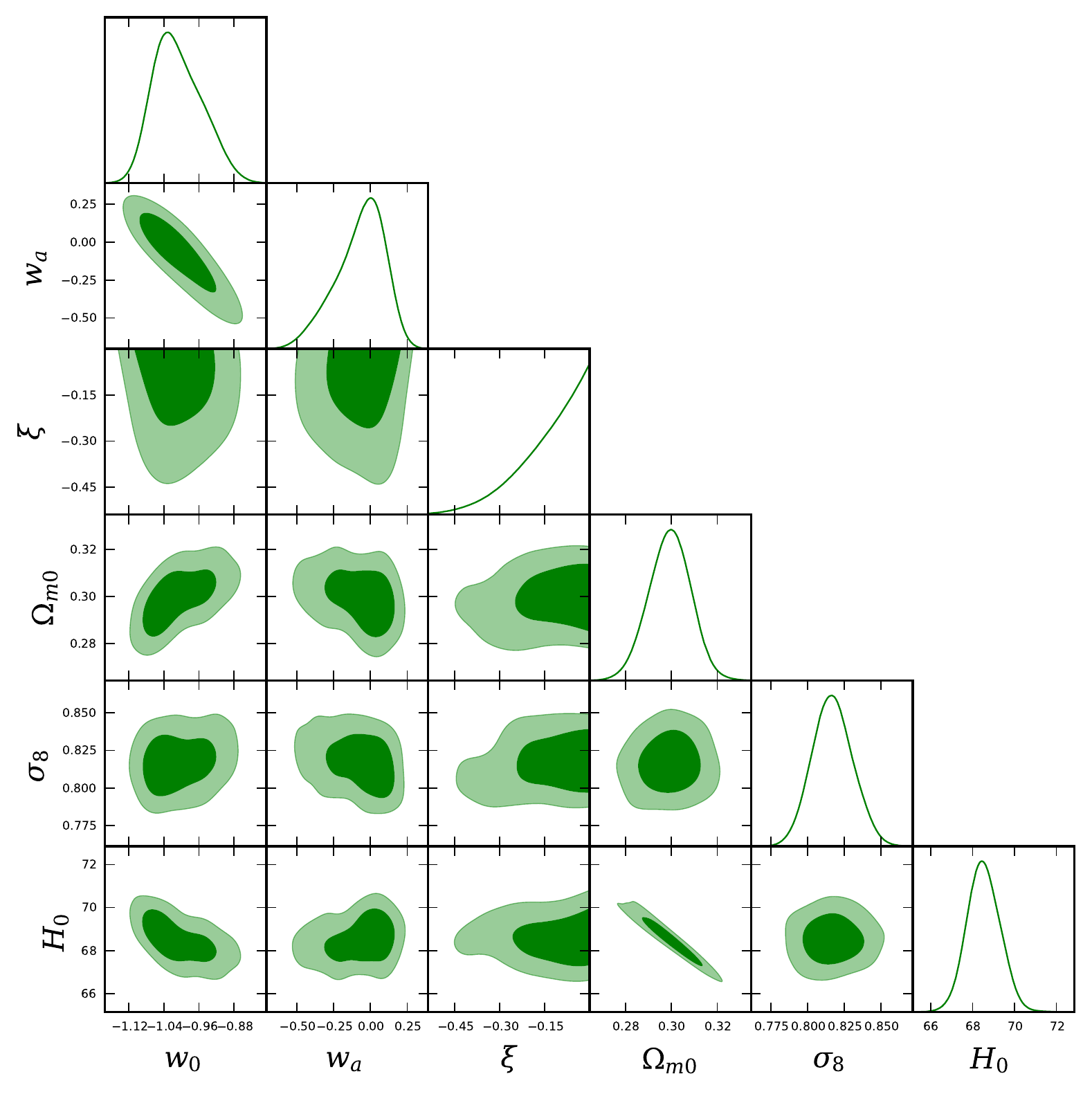}
		\caption{68.3\% and 95.4\% confidence-level contour plots for different combinations of the free parameters of IDE 2 have been shown for the combined analysis Planck TT, TE, EE $+$ lowTEB $+$ BAO $+$ JLA $+$ RSD $+$ WL $+$ CC $+$ R16. }
		\label{fig:contourII}
	\end{figure}

For the interacting DE with dynamical EoS (IDE 2), we present the observational constraints in Table \ref{tab:dynamicalw} using the same combined analysis  Planck TT, TE, EE $+$ lowTEB $+$ BAO $+$ JLA $+$ RSD $+$ WL $+$ CC $+$ R16. The corresponing contour plots at 68.3\%, 95.4\% confidence regions for different combinations of the free parameters of this scenario plus the 1-dimensional posterior distribution of the free parameters have been shown in Figure \ref{fig:contourII}. Our analysis shows that the mode value as well as the best fit values of the present dark energy equation of state has phantom behavior. However, from the observational data, one can infer that, within $1\sigma$ confidence-level, $w_0 = -1$, is compatible. The coupling parameter $\xi$  as seen from Table \ref{tab:dynamicalw} is nonzero which signals for a non-interacting cosmological scenario, while within $1\sigma$ confidence level, $\xi = 0$, is also consistent with the present observational data. That means within $1\sigma$ confidence-level, the interacting model is indistinguishable from $\Lambda$-cosmology.

\begin{figure}%[tbh]
		\includegraphics[width=0.32\textwidth]{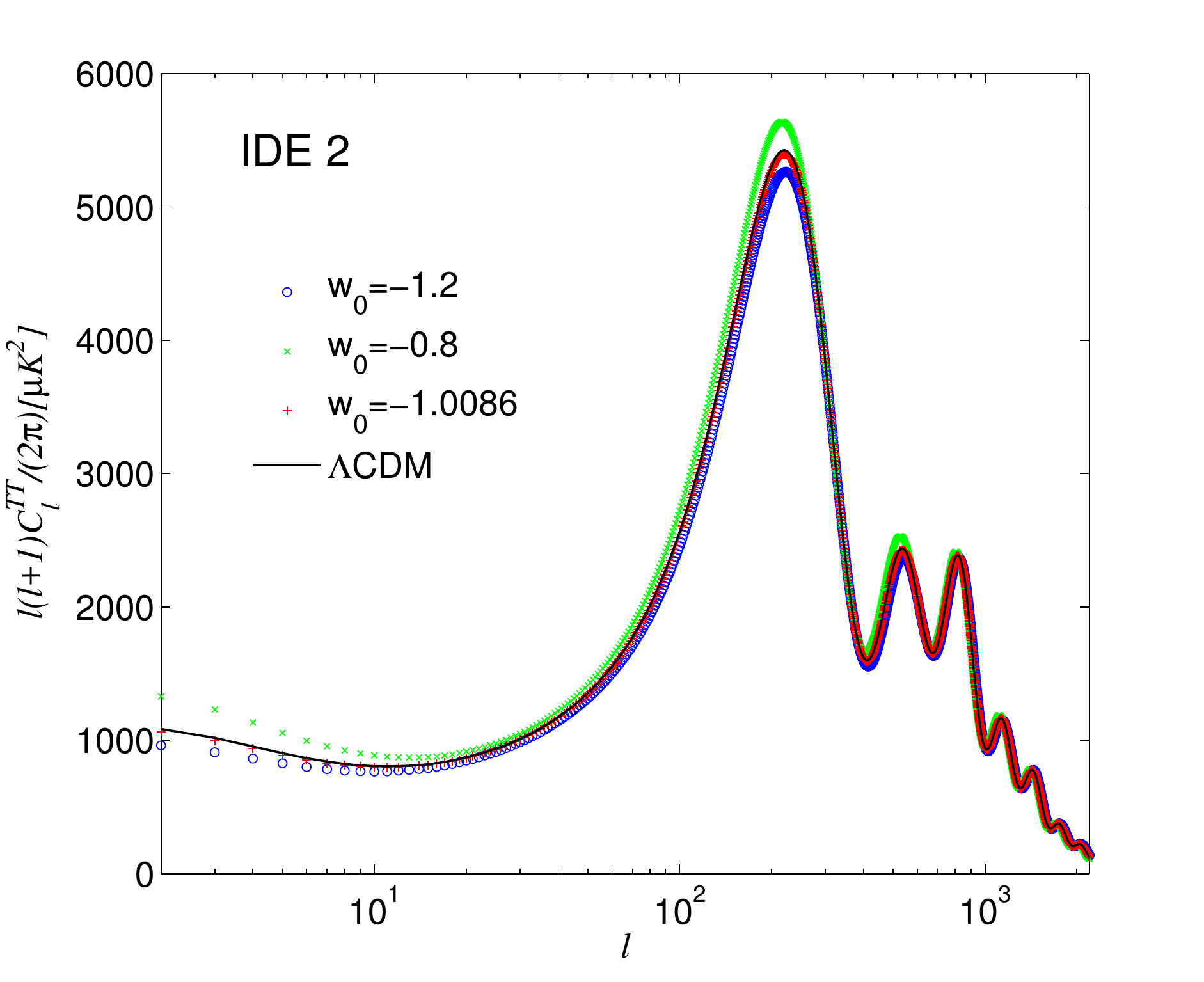}
		\includegraphics[width=0.32\textwidth]{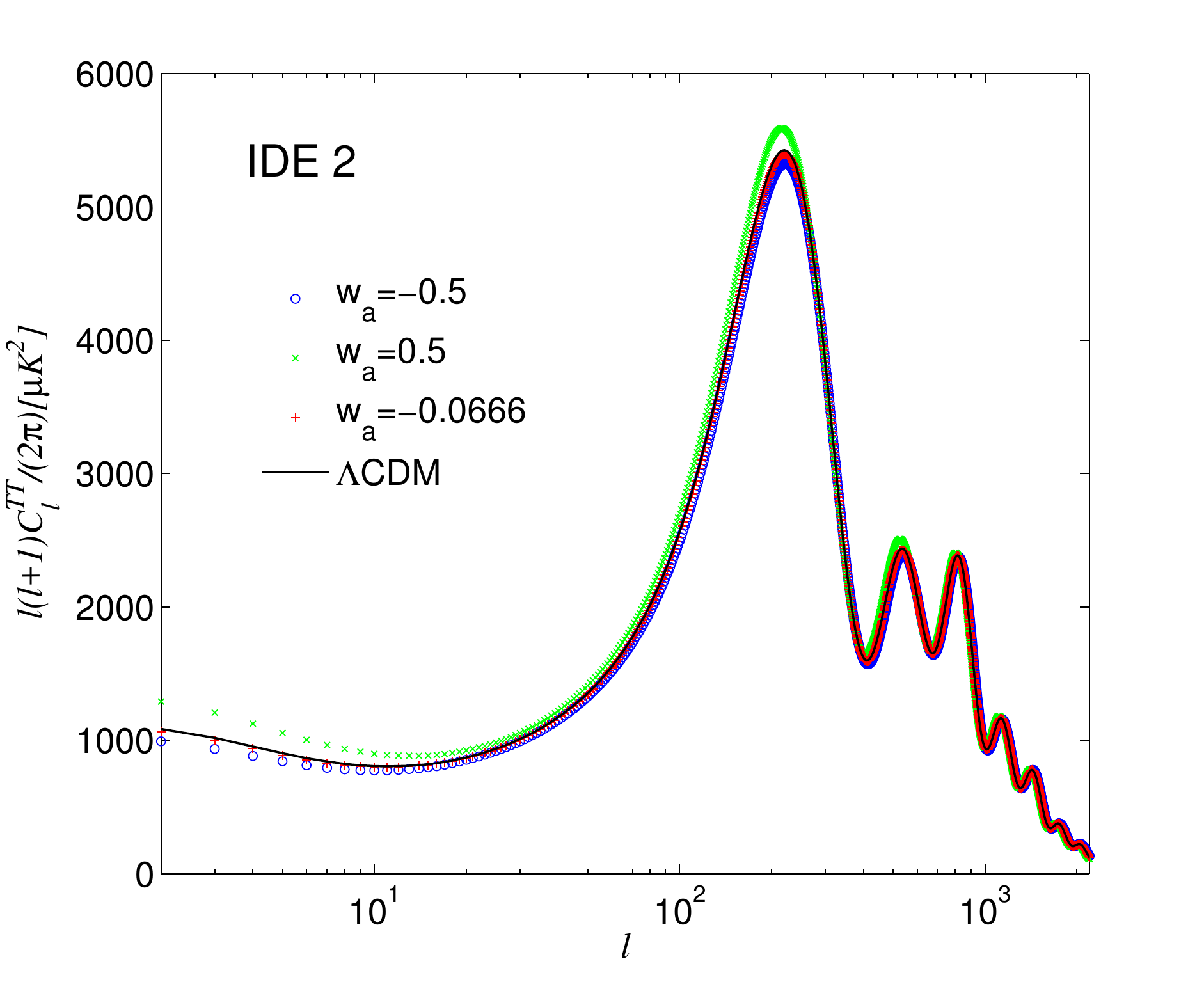}
		\includegraphics[width=0.32\textwidth]{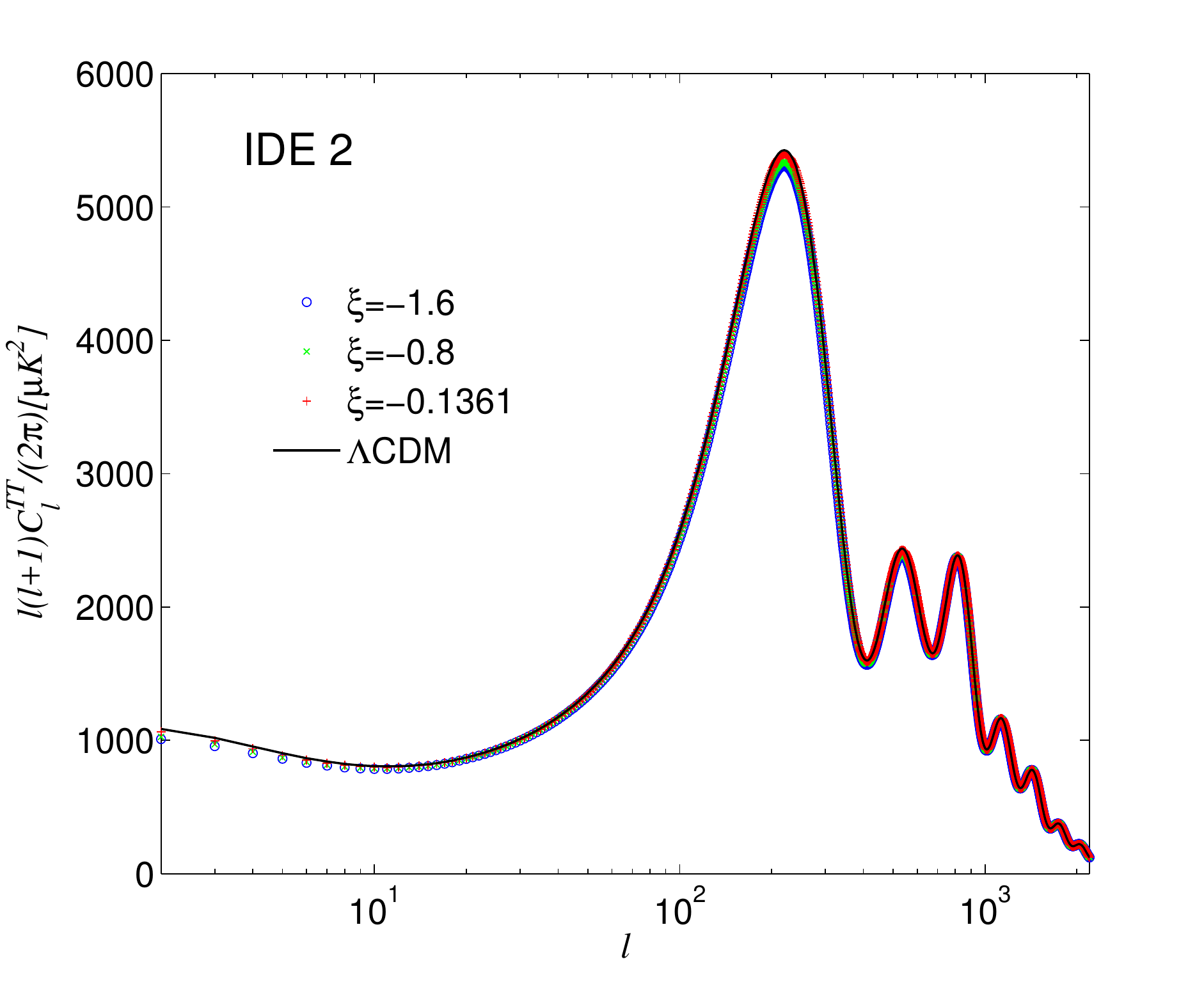}
		\caption{The plots show the angular CMB termperature anisotropy spectra for IDE 2 over the standard $\Lambda$CDM cosmology using the joint analysis data Planck TT, TE, EE $+$ lowTEB $+$ BAO $+$ JLA $+$ RSD $+$ WL $+$ CC $+$ R16. In the left panel we have varied the free parameter $w_0$ (i.e. the current value of the EoS for DE) of the CPL parametrization, in the middle panel we have varied the parameter $w_a$ of the CPL parametrization while the right panel stands for diferent values of the coupling parameter $\xi$. We note that although the curves in the left and middle panels are identified, however, the curves in the right panel are so close to each other such that they are practically indistinguishable from each other. }
		\label{fig:cmbplotII}
	\end{figure}

\begin{figure}%[tbh]
		\includegraphics[width=0.36\textwidth]{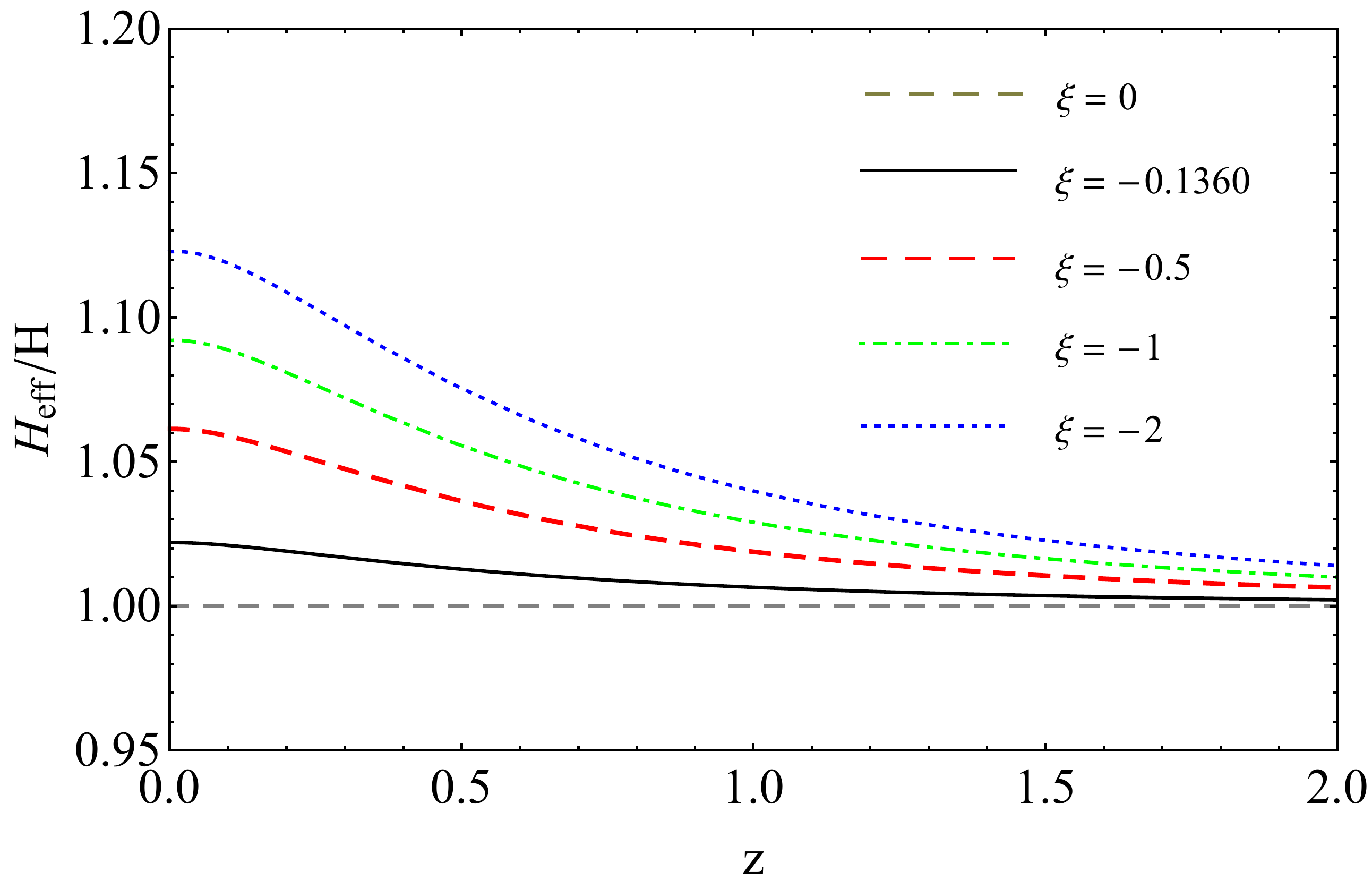}
		\includegraphics[width=0.34\textwidth]{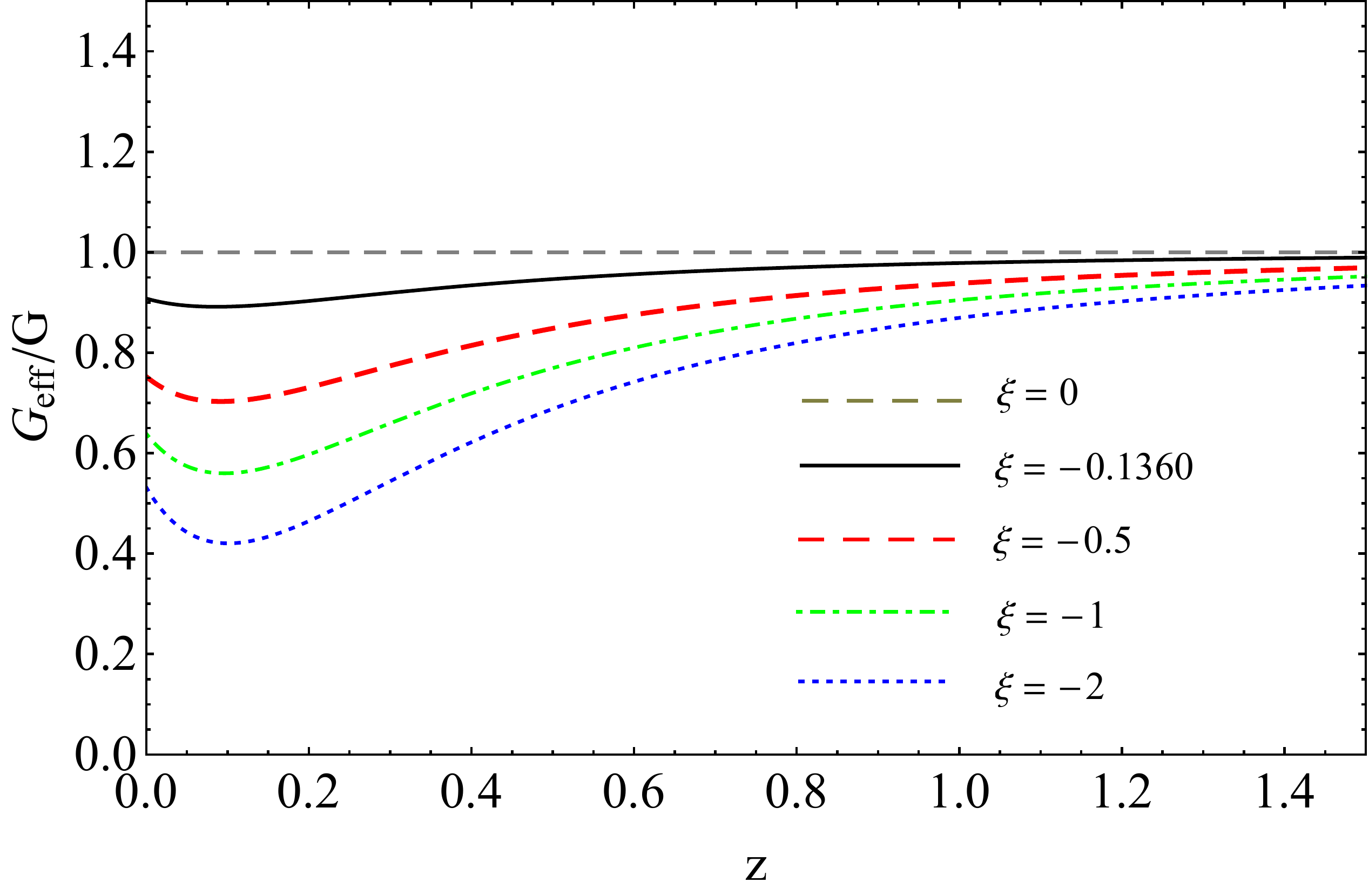}
		\caption{For the interacting dark energy model with its dynamical EoS, $w_x = w_0 + w_a(1 -a)$, in this plot, using different coupling parameters, we show the evolutions of $\mathcal{H}_{eff}/\mathcal{H}$ (left panel) and $G_{eff}/G$ (right panel). The deviation is measured from the non-interacting scenario (i.e. $\xi =0$) and also from the mean value of $\xi = -0.1360$ obtained from the combined observational analysis Planck TT, TE, EE $+$ lowTEB $+$ BAO $+$ JLA $+$ RSD $+$ WL $+$ CC $+$ R16.}
		\label{fig:new2-dynamical}
	\end{figure}

On the other hand, from the CMB spectra (see Figure \ref{fig:cmbplotII}) we observe similar results as we find in IDE 1. That means, the CMB spectra do not show any significant variation for different values of the coupling parameter while for different values of $w_0$, $w_a$, we may expect slight variation on the CMB spectra, for instance, in the left panel of Figure \ref{fig:cmbplotII}, we see that for high quintessential nature of the dark energy equation of state, a slight difference from the $\Lambda$CDM cosmology is observed. Thus, one can assume that if $w_0$ increases, then the deviation would be prominent. Similarly, in the middle panel of Figure \ref{fig:cmbplotII}, we see that for $w_a > 0$, slight variation from $\Lambda$CDM cosmology has appeared. In addition to that, like IDE 1, in Figure \ref{fig:new2-dynamical} we show the evolution of the modified Hubble function $\mathcal{H}_{eff}$ and the modified gravitational constant $G_{eff}$ for different values of the coupling parameter. We observe an equivalent behaviour to that of IDE 1. The plots in Figure \ref{fig:new2-dynamical} clearly  show that as the magnitude of the coupling parameter increases, the deviation of both modified Hubble function and the gravitational constant, increases from that of the non-interacting $w_x$CDM model where $w_x$ is dynamical with the form chosen in this figure. However, one may note that the rate of increment of the quantities $\mathcal{H}_{eff}/\mathcal{H}$ and $G_{eff}/G$ from the non-interacting models as shown in Figure \ref{fig:new2-dynamical}, is slightly lower in compared to the plots in Figure \ref{fig:new1-cons}.

\begin{figure}%[tbh]
		\includegraphics[width=0.38\textwidth]{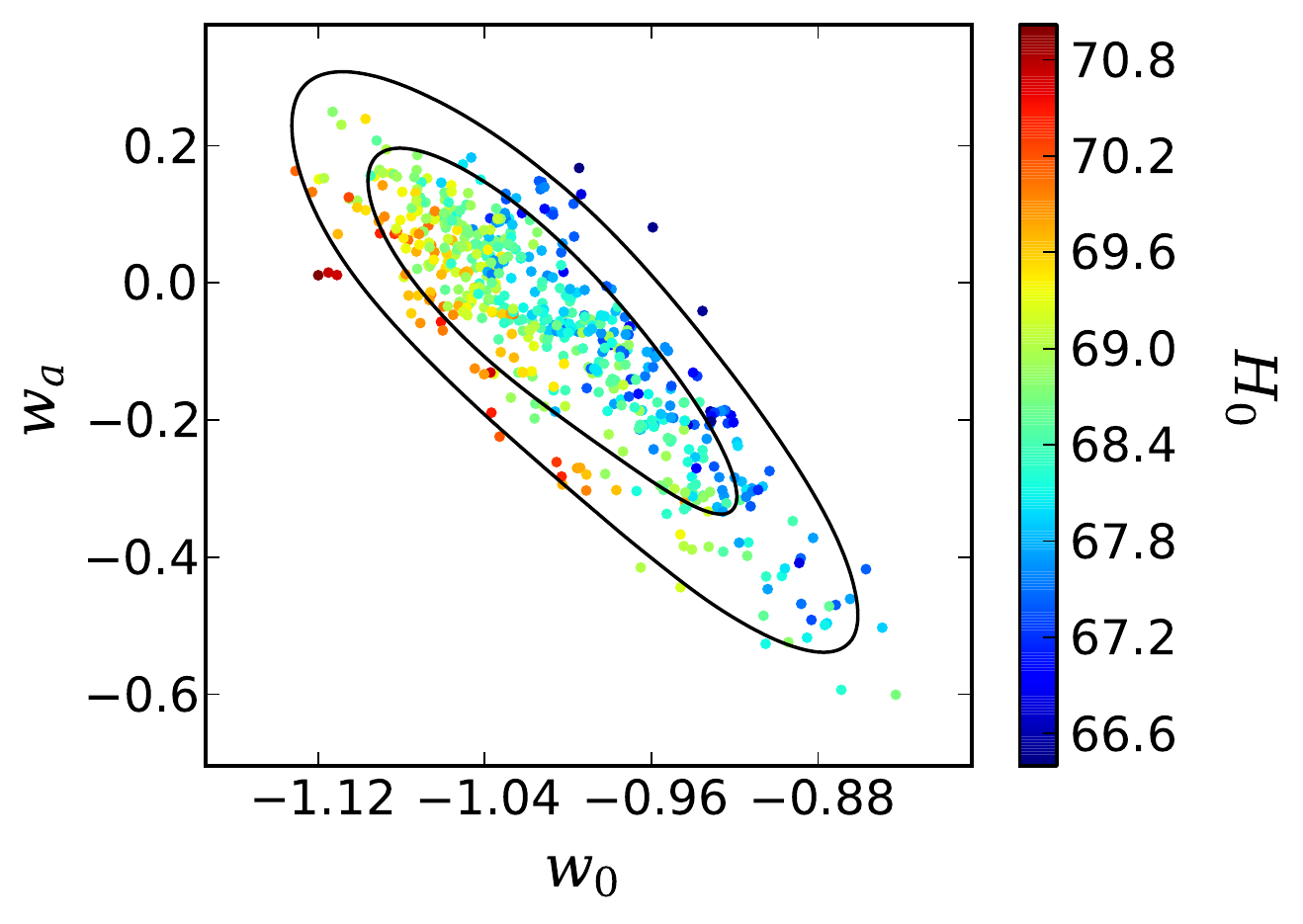}
		\includegraphics[width=0.38\textwidth]{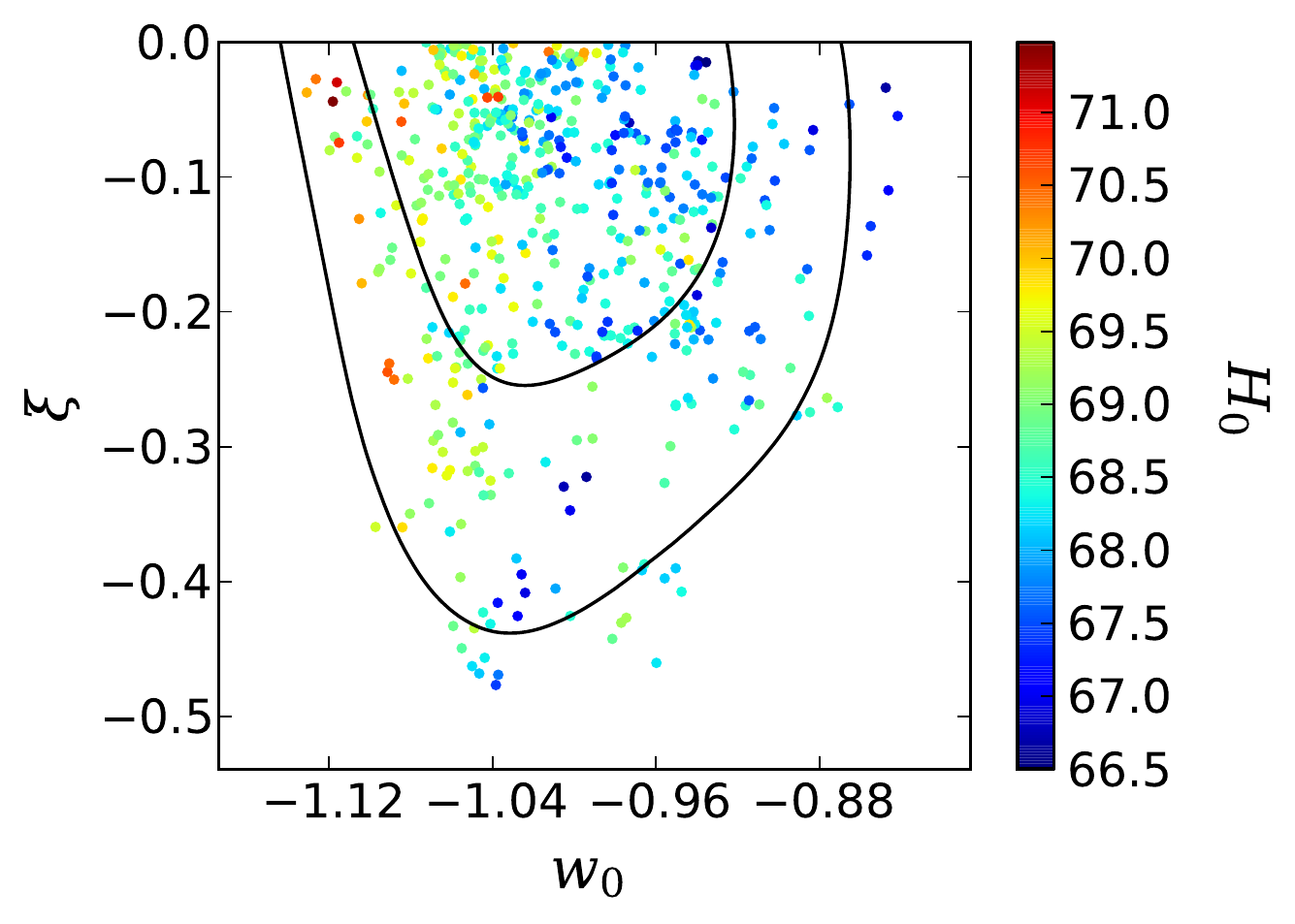}
		\caption{We show the MCMC sample in this figure. In the left panel we show the 2-dimensional posterior distribution for the parameters $(w_0, w_a)$ colored by $H_0$.  In the right panel the sample space contains the $(w_0, \xi)$ plane colored by $H_0$. We note that all the chains have been analyzed with the combined observational data Planck TT, TE, EE $+$ lowTEB $+$ BAO $+$ JLA $+$ RSD $+$ WL $+$ CC $+$ R16. }
		\label{fig:scatter2-IDE2}
	\end{figure}

Using the same combined analysis, in Figure \ref{fig:scatter2-IDE2}, we show the two-dimensional marginalized posterior distributions
for the parameters ($w_a$, $w_0$) and ($w_0$, $\xi$) colored by the $H_0$
sample from the Markov chain Monte Carlo analysis. From the left panel of Figure \ref{fig:scatter2-IDE2} we find that for higher values of $H_0$, the present value of the dark energy equation of state, i.e. $w_0$ favors the phantom behavior while for lower values of $H_0$, the current value of the dark energy equation of state, i.e. $w_0$ shifts toward the quintessence regime. The right panel of Figure \ref{fig:scatter2-IDE2} concludes similar behaviour obtained from its left panel.

%\begin{center}
%\end{center}

%\begin{center}
%\end{center}

%\begin{center}
%\end{center}

%\begin{center}
%\end{center}

%\begin{center}
	\begin{figure}%[tbh]
		\includegraphics[width=0.6\textwidth]{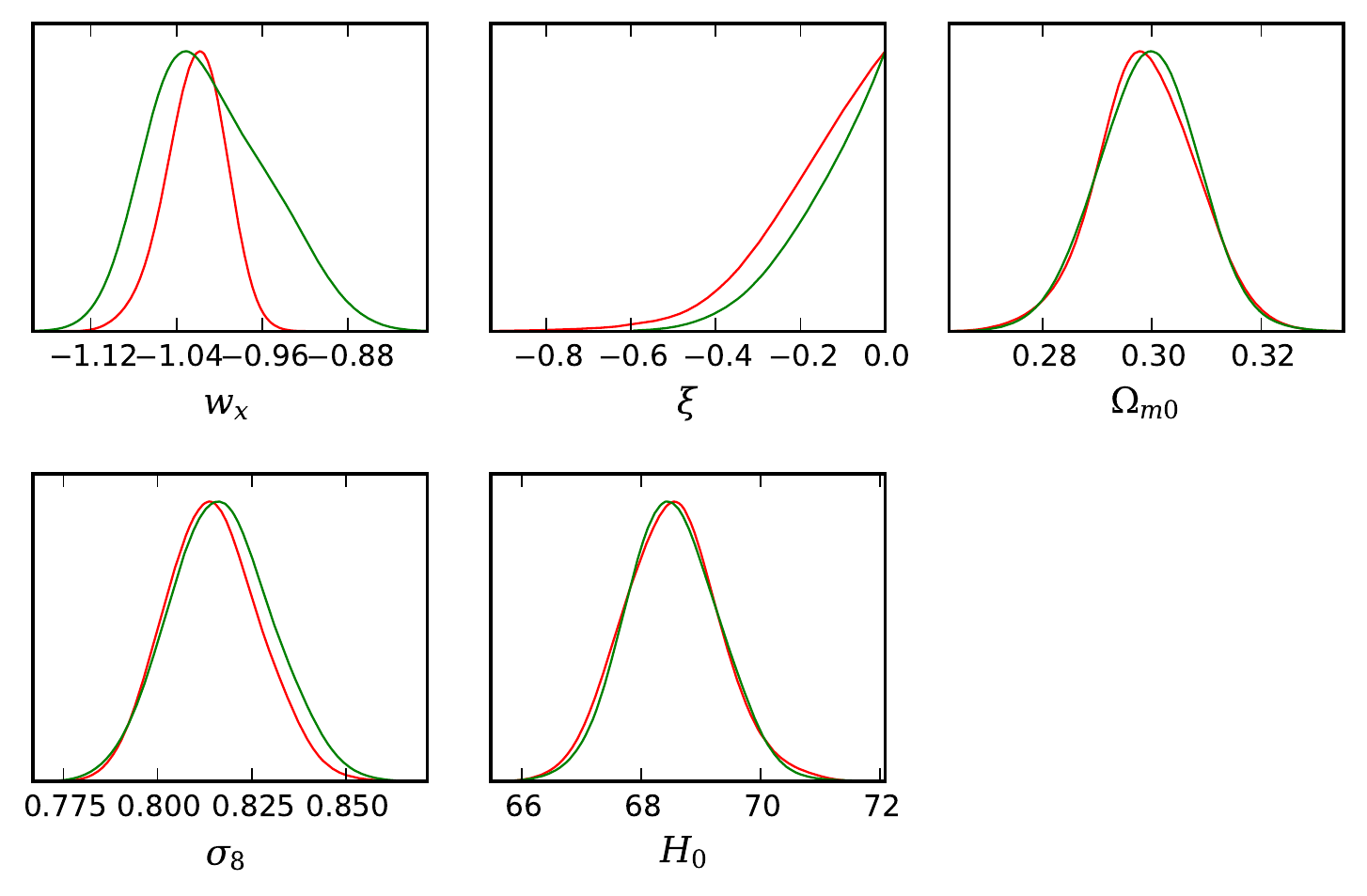}
		\caption{The figure makes a comparison of the one-dimensional posterior distribution for the free and derived parameters for IDE 1 and IDE 2. The observational data used is, Planck TT, TE, EE $+$ lowTEB $+$ BAO $+$ JLA $+$ RSD $+$ WL $+$ CC $+$ R16. We note that in the top left panel, we have used only the parameter $w_x$, which is also the  $w_0$ parameter of the dynamical DE identified by $w_x = w_0 + w_a (1-a)$.}
		\label{fig:posterior-comparison}
	\end{figure}
%\end{center}

%\begin{center}
	\begin{figure}%[tbh]
		\includegraphics[width=0.35\textwidth]{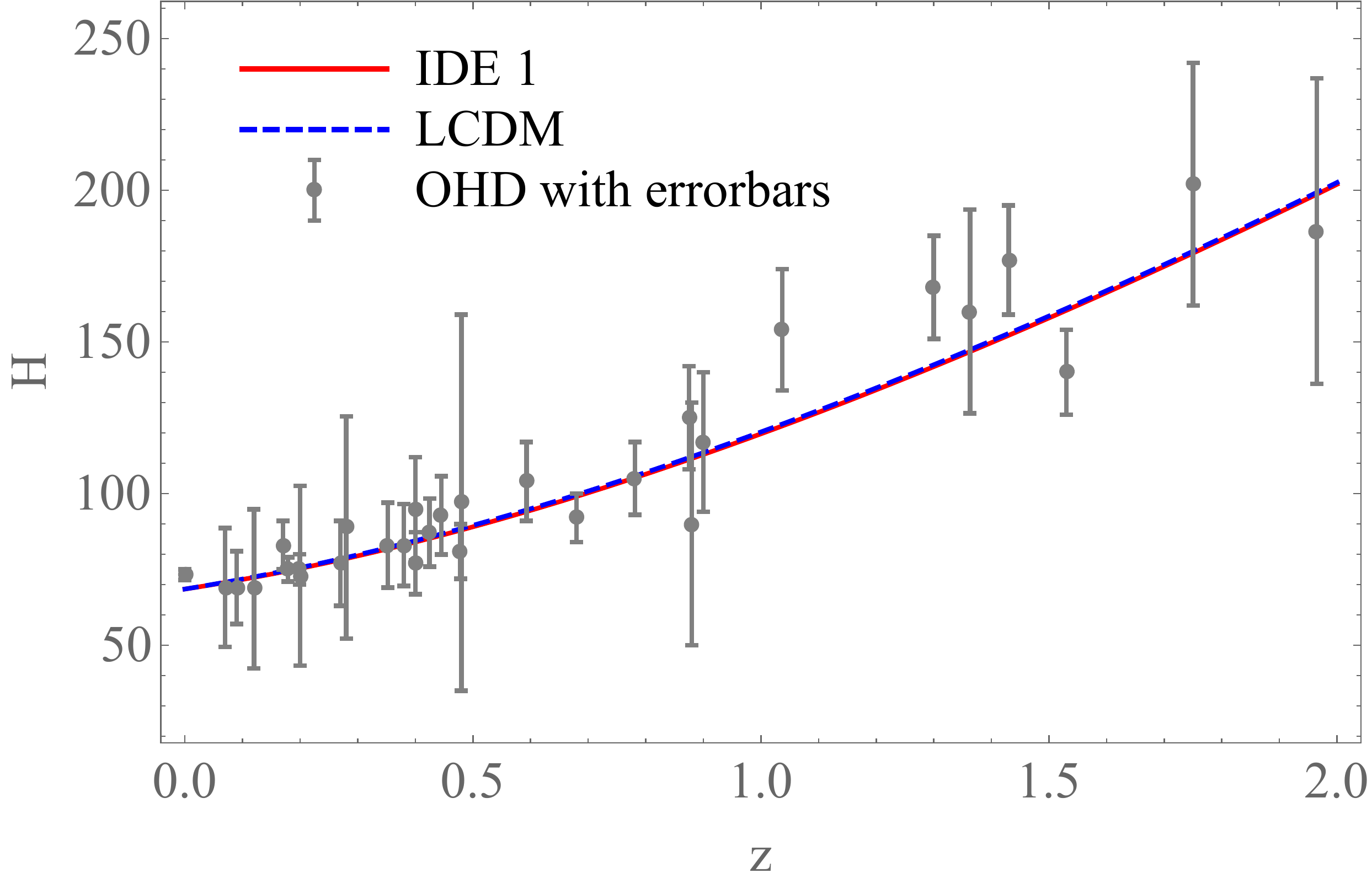}
		\includegraphics[width=0.35\textwidth]{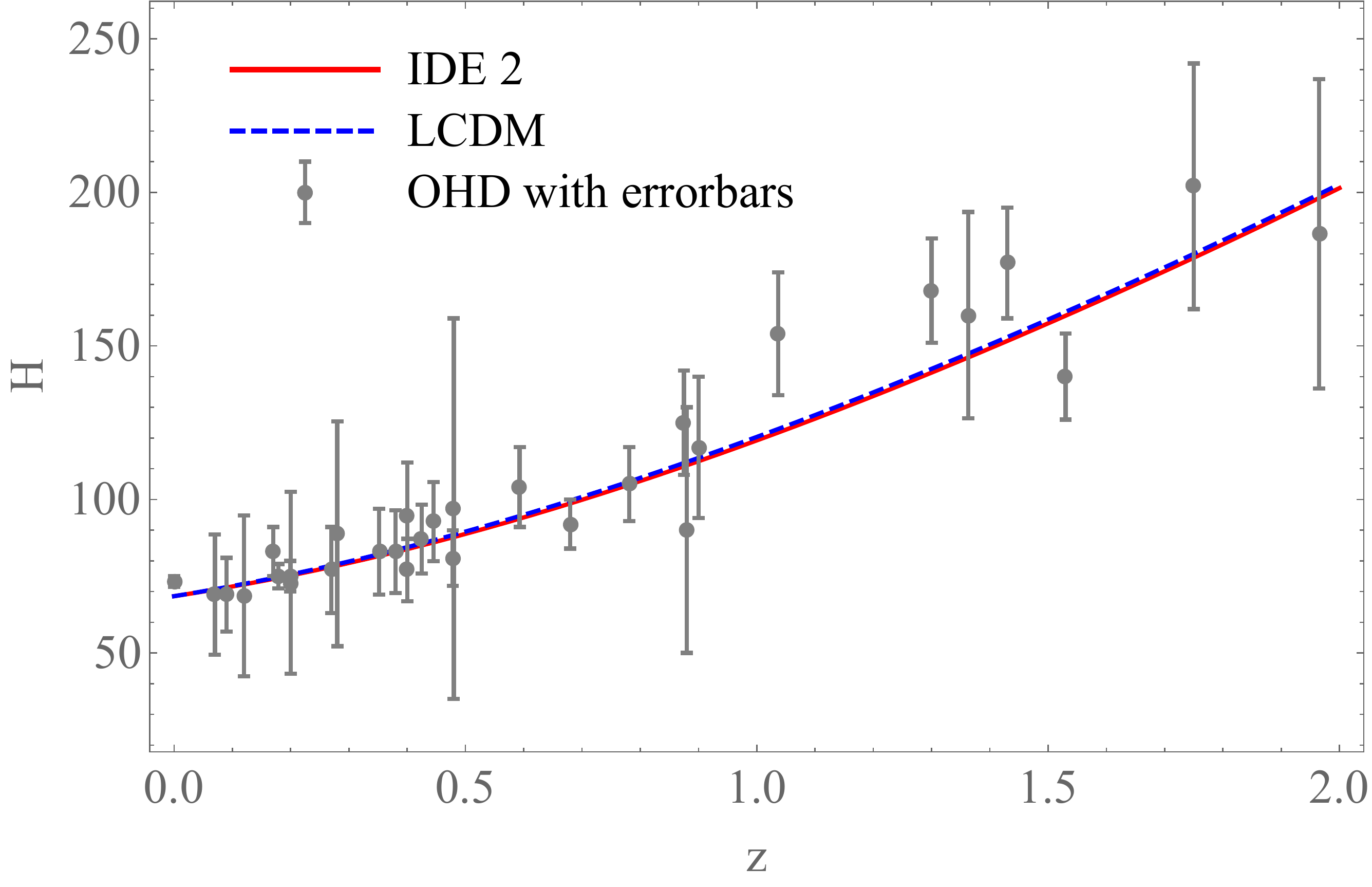}
		\caption{The evolution of the Hubble parameter for two different IDE models in compared to the flat $\Lambda$CDM model and the errorbars of the observed Hubble data are shown. The fitting analysis is same: Planck TT, TE, EE $+$ lowTEB $+$ BAO $+$ JLA $+$ RSD $+$ WL $+$ CC $+$ R16. }

\label{fig:Hubble}
	\end{figure}
	%\end{center}
	
%\begin{center}
	\begin{figure}%[tbh]
		\includegraphics[width=0.38\textwidth]{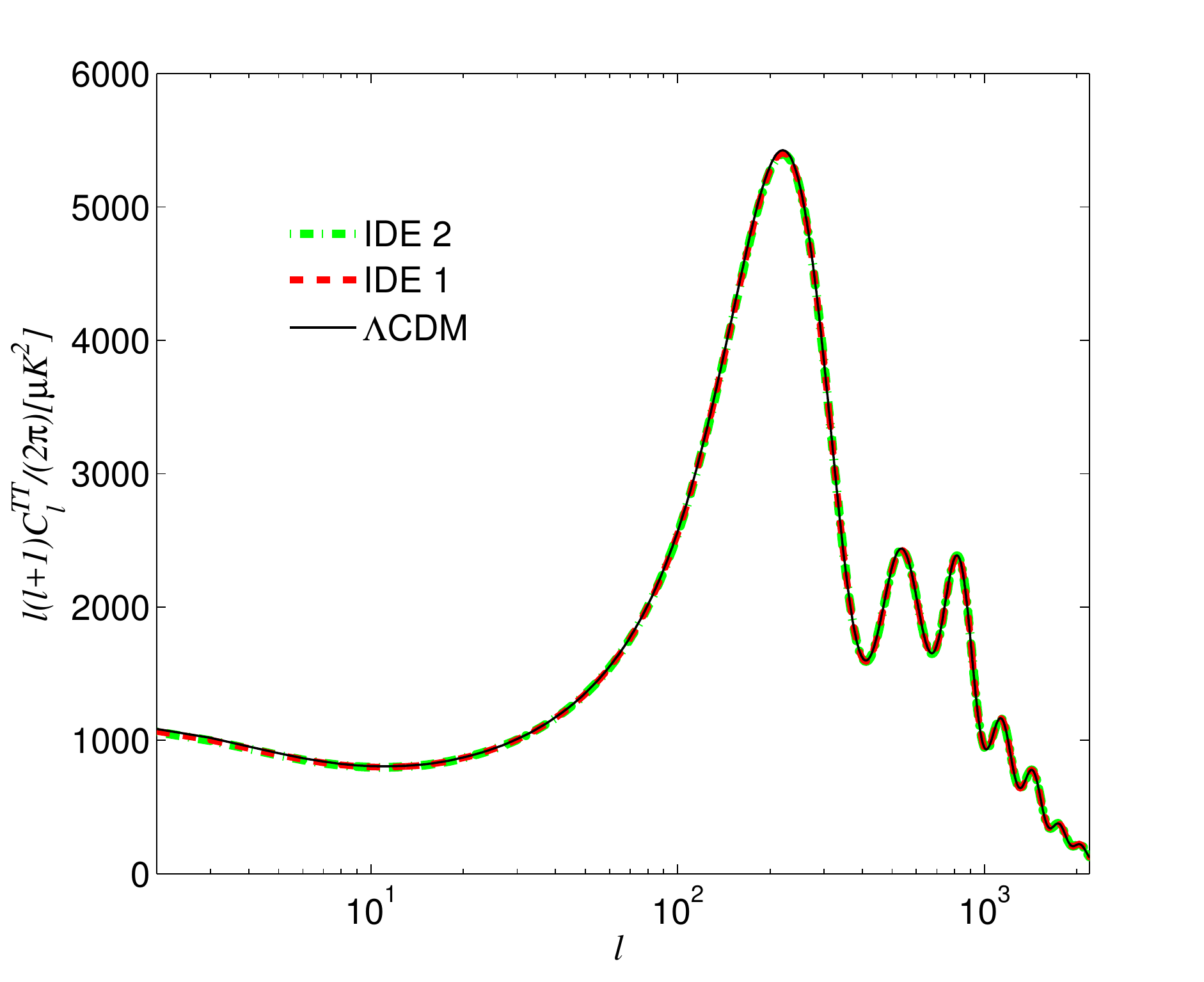}
		\caption{Using the mean values of the free parameters (from the combined analysis Planck TT, TE, EE $+$ lowTEB $+$ BAO $+$ JLA $+$ RSD $+$ WL $+$ CC $+$ R16) associated with IDE 1 and IDE 2, in this plot we show the angular CMB temperature anisotropy spectra to make a comparison of these two interacting scenarios with respect to the $\Lambda$CDM cosmology. The curves as seen from the plot are completely indistinguishable. }
		\label{fig:cmbplot-compare}
	\end{figure}
%\end{center}

%\begin{center}
	\begin{figure}%[tbh]
		\includegraphics[width=0.35\textwidth]{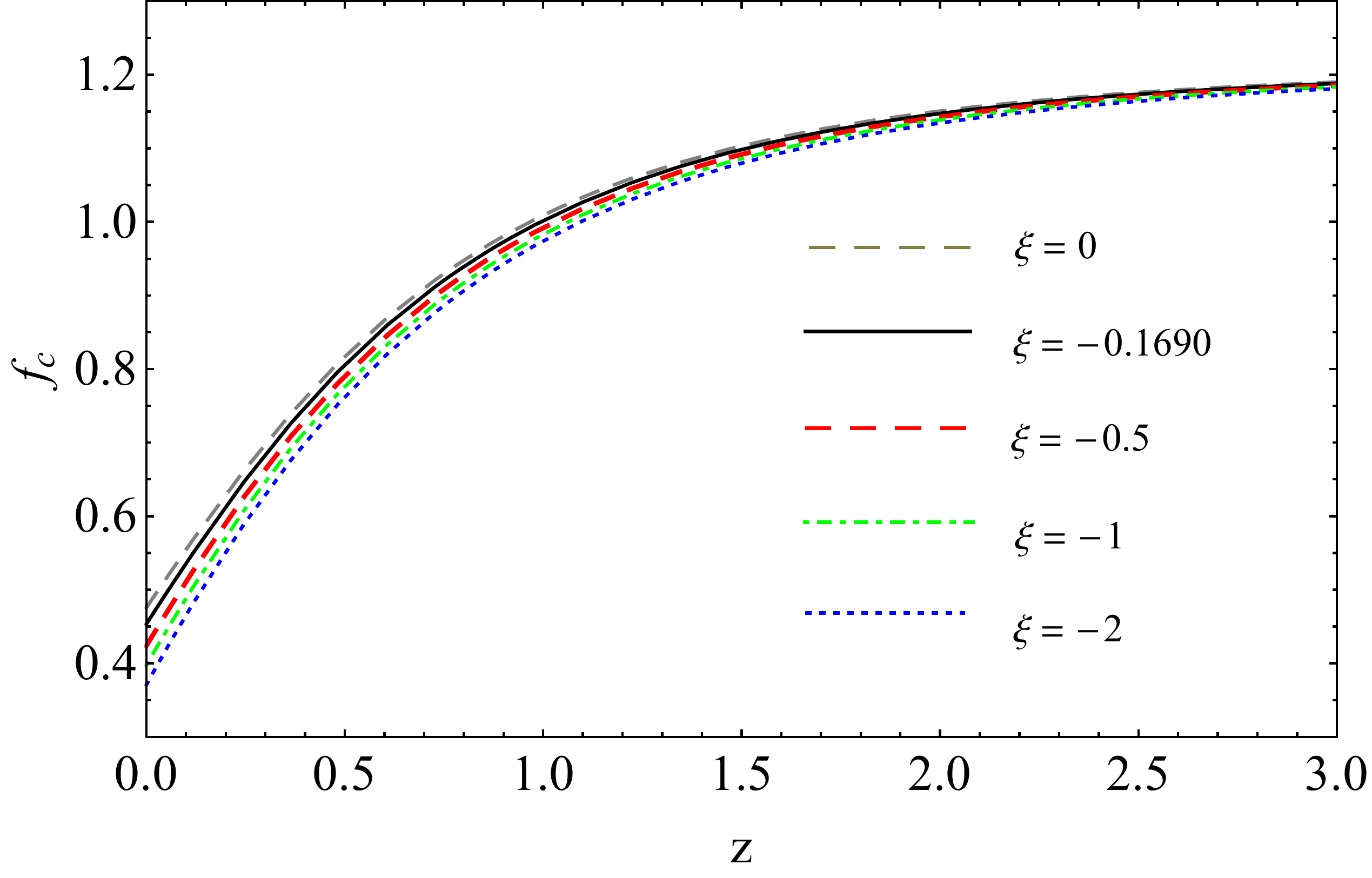}
		\includegraphics[width=0.35\textwidth]{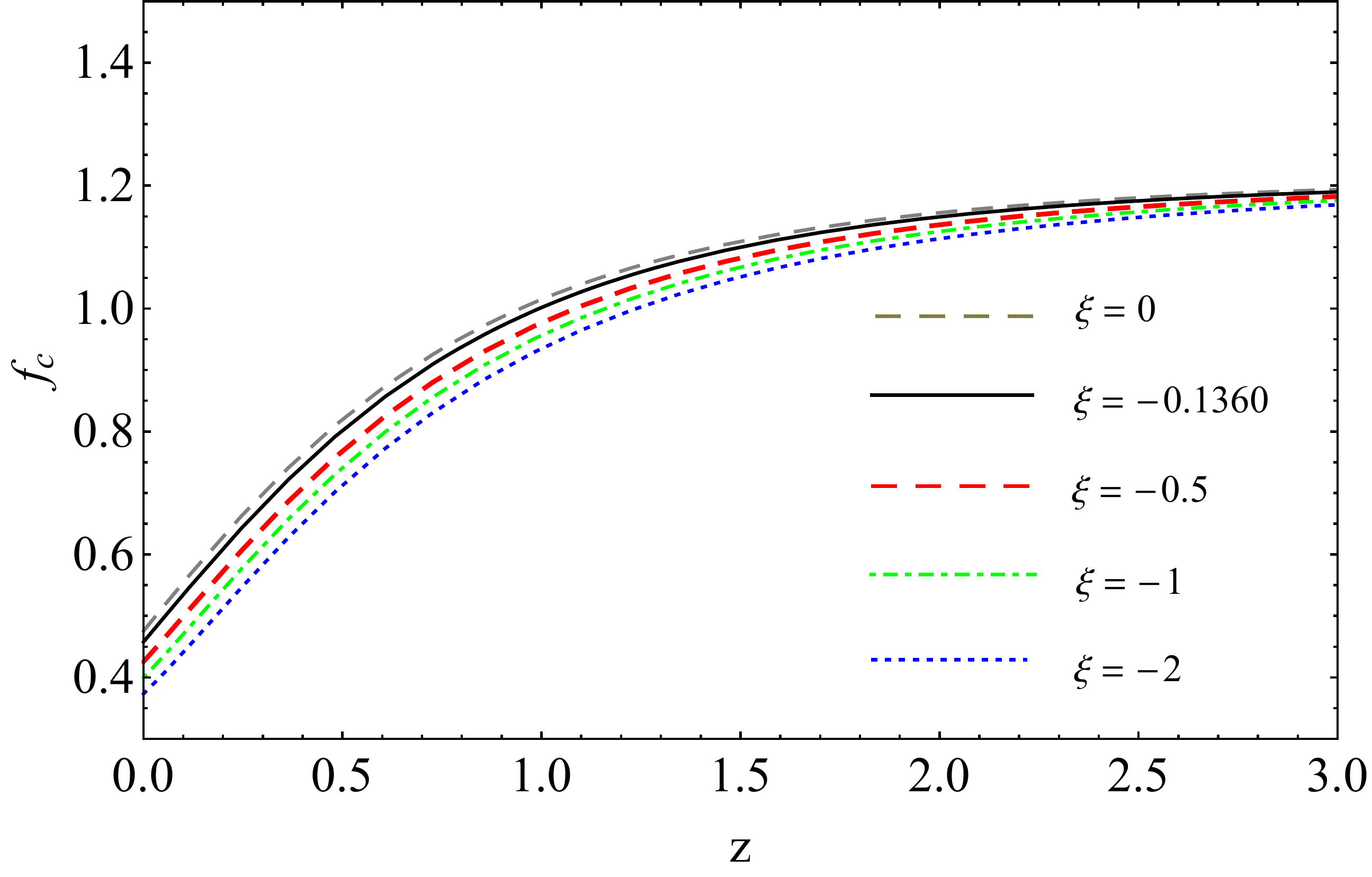}
		\caption{The evolution of growth rate of cold dark matter has been shown for both interacting dark energy models. In the left panel we show the evolution of growth rate of cold dark matter for IDE 1 while the right panel contains the same information but for IDE 2. We use the combined observational data Planck TT, TE, EE $+$ lowTEB $+$ BAO $+$ JLA $+$ RSD $+$ WL $+$ CC $+$ R16. }
		\label{fig:new-growth}
	\end{figure}
%\end{center}

\section{Comparisons between IDE 1 and IDE 2 and with other interactions}
\label{sec-comparison}

Interacting models with dynamical dark energy are quite appealing and interesting. In this work since we consider both constant and time dependent dark energy equation of state, so it is reasonable to measure their qualitative changes both at background and perturbative levels. To compare both the models, in Figure \ref{fig:posterior-comparison} we show the one-dimensional posterior distribution for the free and derived parameters of the interacting models. From the figure we do not observe any significant changes in the parameters. Although a slight difference appears in $w_x$, but this does not seem to be a notable one. Moroever, in Figure \ref{fig:Hubble} we plot the Hubble parameter functions for both IDE models with CC$+$ R16. From this figure we see that both models are very close to the $\Lambda$-cosmology and in fact at low redshift, they are well consistent with the observed data which are shown in the above plots with their error bars. Therereore, it is clear that at background level, both IDE models are very close to each other and they have considerable closure with the $\Lambda$CDM cosmology. It is even very interesting to see that at the perturbative level, we do have the same feature.

If we look at the CMB tempertaure anisotropy spectra in Figure \ref{fig:cmbplot-compare} we see that both IDE 1 and IDE 2 are very close to each other and they are also very close to $\Lambda$CDM. The slight differences between these models can be detected when the growth rate of cold dark matter is encountered into the picture. We illustrate this slight differences between these two interacting models in Figure \ref{fig:new-growth}. In the left panel of Figure \ref{fig:new-growth} we display the growth rate of dark matter when the interacting dark energy has constant EoS where for differnet coupling parameter $\xi$, we show  a couple of plots. From the plots we see that, as $|\xi|$ increases, the growth rate for cold dark matter decreases. The similar behaviour is observed when the interacting dark energy has dynamical character. However, for the dynamical DE,  we have something more. Here, the growth rate for cold dark matter decreases much than growth rate for cold dark matter in the framework of ineracting DE with constant EoS. However, in summary, one can easily conclude that the models IDE 1 and IDE 2 are statistically very close to each other.

\begin{table}
\begin{center}
\caption{\textit{The table displays some known interaction models and the current model with their testable regions.} }
\label{table:coupling}
\begin{tabular}{cccc}
\hline
\hline
Models &  Coupling  & Large-scale stability\\
\hline
Model I    &  $Q= 3 H \xi \rho_{c}$~~~~~& either $w_x\geq -1$ and $\xi \geq 0$ or $w_x\leq -1$ and $\xi \leq 0$\\
Model II   &  $Q= 3 H \xi \rho_{x}$~~~~~& either $w_x\geq -1$ and $\xi \geq 0$ or $w_x\leq -1$ and $\xi \leq 0$\\
Model III  &  $Q = 3 H \xi (\rho_{c}+\rho_{x})$~~~~~& either $w_x\geq -1$ and $\xi \geq 0$ or $w_x\leq -1$ and $\xi \leq 0$\\
Present model & $Q = \xi \dot{\rho}_x$~~~~~& for all $w_x$ and $\xi \leq 0$\\
\hline
\end{tabular}
\end{center}
\end{table}

Moreover, we find that the current interaction model is qualitatively different with other interaction models. As of now, in the current literature, different coupled dark energy models have been introduced and analyzed with the astronomical observations. The most studied and well known coupled DE models include the interactions $Q = 3 H \xi \rho_c$, $Q = 3H \xi \rho_x$, $Q = 3 H \xi (\rho_c +\rho_x)$ ($\protect \xi$ is the coupling parameter) and there are many more, see \cite{Billyard:2000bh, Olivares:2005tb,delCampo:2008jx,Amendola, Koivisto, delCampo:2008sr,Chimento:2009hj, Quartin:2008px, Valiviita:2009nu, Clemson:2011an, Pan:2013rha, Yang:2014hea, Faraoni:2014vra, Yang:2014gza, Nunes:2014qoa, yang:2014vza,thor,barrow, amendola, llinares, Pan:2014afa, Chen:2011cy, Tamanini:2015iia, Pan:2012ki, Duniya:2015nva, Valiviita:2015dfa, Yang:2016evp, Pan:2016ngu, Mukherjee:2016shl, Sola:2016ecz, Sharov:2017iue,Cai:2017yww, Santos:2017bqm, Mifsud:2017fsy, Salvatelli:2014zta, Nunes:2016dlj,Kumar:2016zpg,vandeBruck:2016hpz, Yang:2017yme, Kumar:2017dnp}. In this work we introduce an interaction with the form $Q \propto \dot{\rho}_x$ which is different from the others since the variation of the dark energy density is considered here. 
This interaction has a great advantage in compared to some well known interaction models. To clarify such an issue, let us recall the observational
bounds on the existing interacting models in the literature.
The past analysis of coupled dark energy models in the large scale universe reports that the models with $Q = 3 H \xi \rho_c$, $Q = 3H \xi \rho_x$, $Q = 3 H \xi (\rho_c +\rho_x)$, can only be tested for two separate intervals, that means when $(w_x \geq -1,\, \xi \geq 0)$ or $(w_x \leq -1,\, \xi \leq 0)$.  See table \ref{table:coupling} where we present some well known interaction models as well as the present interaction model with their testable regions. Clearly, while constraining the above interaction models we perceive a discontinuity in the parametric space. In order to remove such discrepancy in the interaction models, a recent investigation \cite{ypb} found that, the whole parametric space can be tested if a phenomenological factor $(1+w_x)$ is introduced in  the interaction function, $Q$. Originally, the hints to introduce such factor $(1+w_x)$ into the interaction function is motivated from the pressure perturbation of dark energy. Using such formalism, three interactions have been tested with positive results, namely, $Q= 3 H \xi (1+w_x)\rho_x$, $Q = 3 H \xi (1+w_x) \rho_c \rho_x/(\rho_c+\rho_x)$, $Q=  3 H \xi (1+w_x) \rho_x^2/\rho_c$ in \cite{ypb}. However, the current interaction provides something more.  We see that the interaction (\ref{Q-1}) does not need any extra factor $(1+w_x)$ from the outside to test the entire parametric space since it automatically contains such factor within it. Thus, this interaction differs with other interaction models in this way and it is worth noting that the entire parameteters space can be constrained unlike other interaction models.

\section{The tension on $H_0$: Alleviation through interacting dark energy}
\label{sec-tension}

From the analysis of cosmic microwave background temperature
and polarization data by Planck \cite{Ade:2015xua}, the Hubble
constant value is constrained to be $H_0= 67.27 \pm 0.66$ km~s$^{-1}$~Mpc$^{-1}$,
assuming $\Lambda$CDM as the base model while on the other hand, from the local
measurements performed recently by Riess et al. \cite{Riess:2016jrr}, the Hubble
constant appears to be $H_0 = 73.24 \pm 1.74$ km~s$^{-1}$~Mpc$^{-1}$. This local
Hubble constant value is about $3\sigma$ higher than its prediction by the
Planck's measurements. The difference appearing in the estimation of $H_0$ from the local and global measurements is quite large.
It is indeed a very big issue\footnote{The continuous
progress towards more consistent observational measurements could reveal
some more information in this regard, see for instance \cite{Lin:2017ikq,  Lin:2017bhs}.}
in the cosmological theory $-$ known as the tension on $H_0$!

The interacting dark energy mechanism carries a reasonable justification
to reconcile such tension that has already been suggested in some
recent works \cite{Kumar:2016zpg, Kumar:2017dnp, DiValentino:2017iww} where
in \cite{DiValentino:2017iww} the dark energy equation of state
was constant and allowed to be phantom while in the other two works, namely,
in Refs. \cite{Kumar:2016zpg} and \cite{Kumar:2017dnp}, the dark energy equation of state
was respectively constant (other than the cosmological constant) and the cosmological constant.

The current interaction model is completely different in
compared to the works \cite{Kumar:2016zpg, Kumar:2017dnp, DiValentino:2017iww}
because here the model can be tested without any restriction on the
dark energy equation of state as explained earlier. Additionally,
here we investigate the dynamics of the universe for both constant and
dynamical dark energy equation of state. We also note that for both the models,
the combined analysis has been fixed to be Planck TT, TE, EE $+$
lowTEB $+$ BAO $+$ JLA $+$ RSD $+$ WL $+$ CC $+$ R16.

\begin{itemize}

\item \textbf{Constant EoS in DE and $H_0$ tension:} We investigate the tension on
$H_0$ for four different intervals of the dark energy equations of state: (i) we allow a wide range of the dark energy equation of state from quintessence to phantom, in particular, $w_x \in [-1.5, -0.9]$, (ii) we consider the dark energy equation of state to vary from the cosmological constant bundary to its beyond, hence, we fix $w_x \in [-2, -1]$,  (iii) we consider only phantom dark energy with
when $w_x \in [-2, -1.01]$, and finally, (iii) we consider a super phantom dark energy equation of state as $w_x \in [-2, -1.2]$. For four different priors on the dark energy equation of state, we present the astronomical constraints on the cosmological parameters in Table \ref{tab:tension1} where the mean values of the parameters are reported. The two dimensional contour plots of the parameters ($w_x$, $H_0$) have been shown in Figure \ref{fig:tension1} for  four different priors of the dark energy equation of state. The analyses show that for $w_x \in [-1.5,-0.9]$ and $w_x \in [-2, -1]$, and $w_x \in [-2, -1.01]$, the Hubble constant value is close to the measured value of $H_0$ ($= 67.27 \pm 0.66$ km~s$^{-1}$~Mpc$^{-1})$ from Planck \cite{Ade:2015xua} and the tension is released at $< 2 \sigma$ CL. While on the other hand, for $w_x \in [-2, -1.2]$, the result is slightly different because here $H_0$ catches the local Hubble constant value measured by Riess et al. ($H_0 = 73.24 \pm 1.74$ km~s$^{-1}$~Mpc$^{-1}$) \cite{Riess:2016jrr} within the confidence level slightly greater than $2\sigma$. So, the tension is not released here. Thus, we see that for the current interaction model, the tension can be released. However, one can note that if the dark energy has a strong phantom nature, for instance here $w_x < -1.2$, then some discrepancies may appear. In summary, the phantom dark energy equation of state might resolve the current tesnion on $H_0$ in agreement with a recent work \cite{DiValentino:2017iww}, but however, the physics with high phantom dark energy equation of state needs further investigations. Finally, we remark that we are dealing with phenomenological models of interaction!

%\begin{center}
\begin{table}%[tbh]
\caption{For the constant dark energy
equation of state, $w_x$, we present the results of the interaction model for
four different regions of $w_x$. The analysis has been performed for
the combined analysis Planck TT, TE, EE $+$ lowTEB $+$ BAO $+$ JLA $+$ RSD $+$ WL $+$ CC $+$ R16. As mentioned earlier, $\Omega_{m0}= \Omega_{c0}+\Omega_{b0}$.  }
\begin{tabular}{cccccccc}
\hline\hline
Parameters & $w_x \in [-1.5, -0.9]$ &  $w_x \in [-2, -1]$ & $w_x \in [-2, -1.01]$ &  $w_x \in [-2, -1.2]$
\\ \hline
$\Omega_ch^2$ & $ 0.11694_{-    0.00128-    0.00353}^{+    0.00194+    0.00302}$ & $    0.11702_{-    0.00129-    0.00311}^{+    0.00161+    0.00287}$ & $0.11747_{-    0.00132-    0.00258}^{+    0.00136+    0.00264}$ & $    0.12011_{-    0.00121-    0.00376}^{+    0.00206+    0.00332}$\\

$\Omega_bh^2$  &  $ 0.02230_{-    0.00014-    0.00028}^{+    0.00014+    0.00029}$ & $0.02228_{-    0.00014-    0.00028}^{+    0.00015+    0.00028}$ & $0.02226_{- 0.00014-    0.00029}^{+    0.00015+    0.00029}$ & $ 0.02202_{-    0.00012-    0.00025}^{+    0.00013+    0.00026}$\\

$100\theta_{MC}$ & $ 1.04070_{- 0.00033-    0.00060}^{+    0.00032+    0.00062}$ & $    1.04069_{-    0.00027-    0.00059}^{+ 0.00029+ 0.00059}$ & $1.04066_{-    0.00030-    0.00064}^{+    0.00036+    0.00060}$ &$ 1.04023_{- 0.00030-    0.00064}^{+    0.00031+    0.00061}$\\

$\tau$ &  $    0.06765_{-    0.01862-    0.03345}^{+    0.01602+    0.03418}$ & $    0.06539_{-    0.01700-    0.03331}^{+    0.01775+    0.03354}$ & $0.06178_{-    0.01627-    0.03299}^{+    0.01613+    0.03366}$ &$    0.03222_{-    0.01690-    0.02221}^{+    0.01031+    0.02317}$\\

$n_s$ &  $    0.97639_{-    0.00428-    0.00822}^{+    0.00419+    0.00870}$ & $    0.97559_{-    0.00390-    0.00784}^{+    0.00385+    0.00842}$ & $0.97491_{-    0.00366-    0.00804}^{+    0.00416+    0.00731}$ &$    0.96665_{-    0.00368-    0.00674}^{+    0.00339+    0.00693}$\\

$\mathrm{ln}(10^{10}A_s)$ &  $    3.07394_{-    0.03451-    0.06387}^{+    0.03206+    0.06496}$ & $    3.07116_{-    0.03289-    0.06678}^{+    0.03186+    0.06265}$ & $3.06402_{-    0.03180-    0.06427}^{+    0.03184+    0.06524}$ &$3.01185_{-    0.03145-    0.04863}^{+    0.02230+    0.05185}$\\

$w_x$ &  $   -1.02320_{-    0.02383-    0.05423}^{+    0.03135+    0.05197}$ & $   -1.03245_{-    0.00939-    0.04405}^{+    0.03058+    0.03245}$ & $-1.04669_{-    0.01327-    0.04508}^{+    0.03195+    0.03669}$ &$   -1.20660_{-    0.00092-    0.01332}^{+    0.00660+    0.00660}$\\

$\xi$ & $   -0.23758_{-    0.08064-    0.28867}^{+    0.23758+    0.23758}$ & $   -0.20708_{-    0.03706-    0.38948}^{+    0.20708+    0.20708}$ & $-0.09160_{-    0.03856-    0.11641}^{+    0.09160+    0.09160}$ &$   -0.02830_{-    0.00377-    0.05692}^{+    0.02830+    0.02830}$\\

$\Omega_{m0}$ & $    0.29786_{-    0.00930-    0.01803}^{+    0.00969+    0.01723}$ & $    0.29640_{-    0.00723-    0.01474}^{+    0.00756+    0.01418}$ & $0.29454_{-    0.00754-    0.01481}^{+    0.00764+    0.01450}$ &$    0.27442_{-    0.00663-    0.01308}^{+    0.00637+    0.01300}$\\

$\sigma_8$ & $    0.81384_{-    0.01288-    0.02521}^{+    0.01299+    0.02568}$ & $    0.81543_{-    0.01290-    0.02615}^{+    0.01284+    0.02542}$ & $0.81797_{-    0.01357-    0.02544}^{+    0.01409+    0.02489}$ & $ 0.84356_{-    0.01284-    0.02679}^{+    0.01297+    0.02624}$\\

$H_0$ &  $   68.54857_{-    0.80198-    1.43037}^{+    0.76687+    1.50829}$ & $   68.72657_{-    0.72678-    1.27043}^{+    0.58200+    1.35008}$ & $69.04942_{-    0.76261-    1.41469}^{+    0.66552+    1.40722}$ &$   72.13901_{-    0.60082-    1.08626}^{+    0.56189+    1.11083}$\\
\hline
\end{tabular}%
\label{tab:tension1}
\end{table}
%\end{center}

\item \textbf{Dynamical EoS in DE and $H_0$ tension:} For dynamical dark energy, we perform same analysis considering four different ranges of $w_0$, the current value of the dark energy equation of state $w_{x} = w_0 + w_a (1-a)$. The constraints on the model parameters have been summarized in Table \ref{tab:tension2} where their mean values have been shown while the contour plots for the parameters ($w_0$, $H_0$) for four different priors on $w_0$ have been shown in Figure \ref{fig:tension2}. We find that the interaction model with dynamical DE preserves almost similar behavior to that of the interaction model with constant dark energy equation of state. For  $w_0 \in [-1.5,-0.9]$, $w_0 \in [-2, -1]$, and $w_0 \in [-2, -1.01]$, the tension on $H_0$ is released within the confidence-level interval ($1\sigma, 2 \sigma$).  On the other hand, for $w_0 \in [-2, -1.2]$, the tension is not released. That means IDE 2 has similarities with IDE 1 as we can see.

%\begin{center}
	\begin{figure}%[tbh]
		\includegraphics[width=0.24\textwidth]{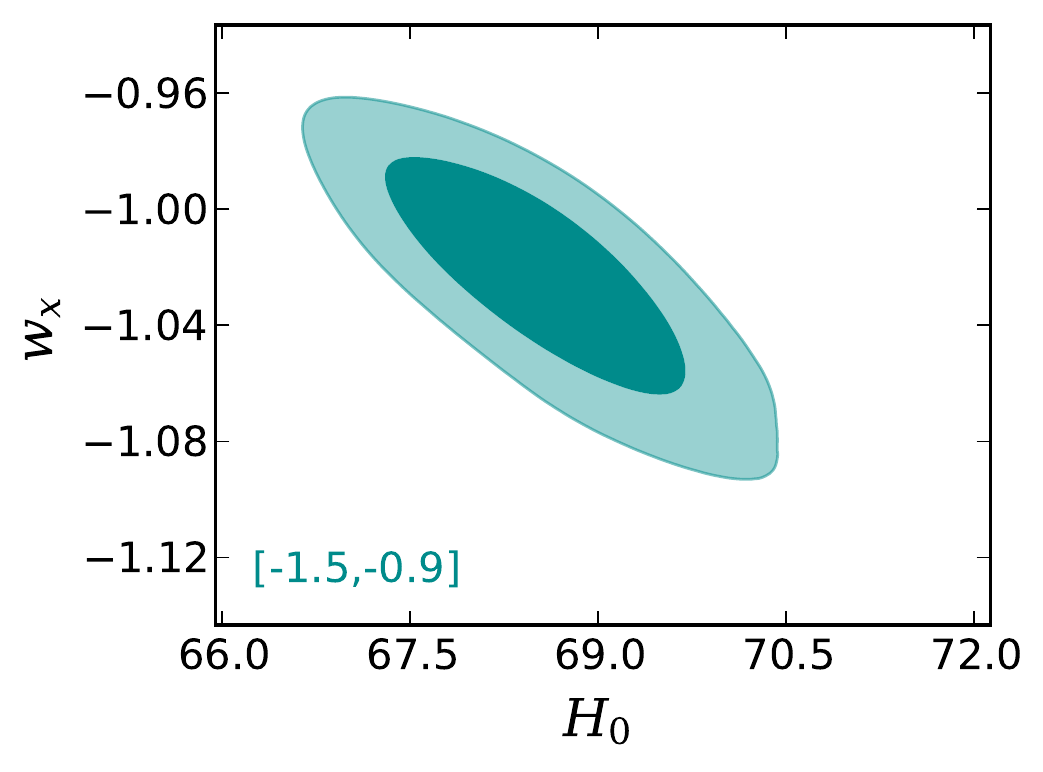}
		\includegraphics[width=0.24\textwidth]{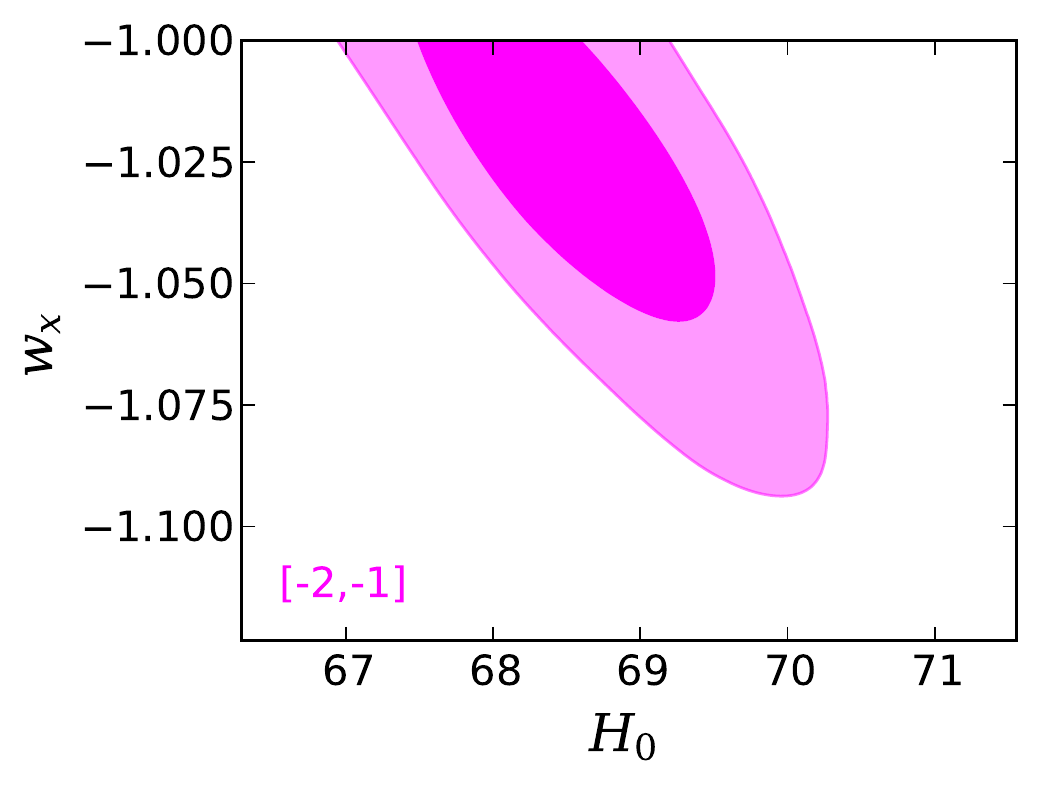}
		\includegraphics[width=0.24\textwidth]{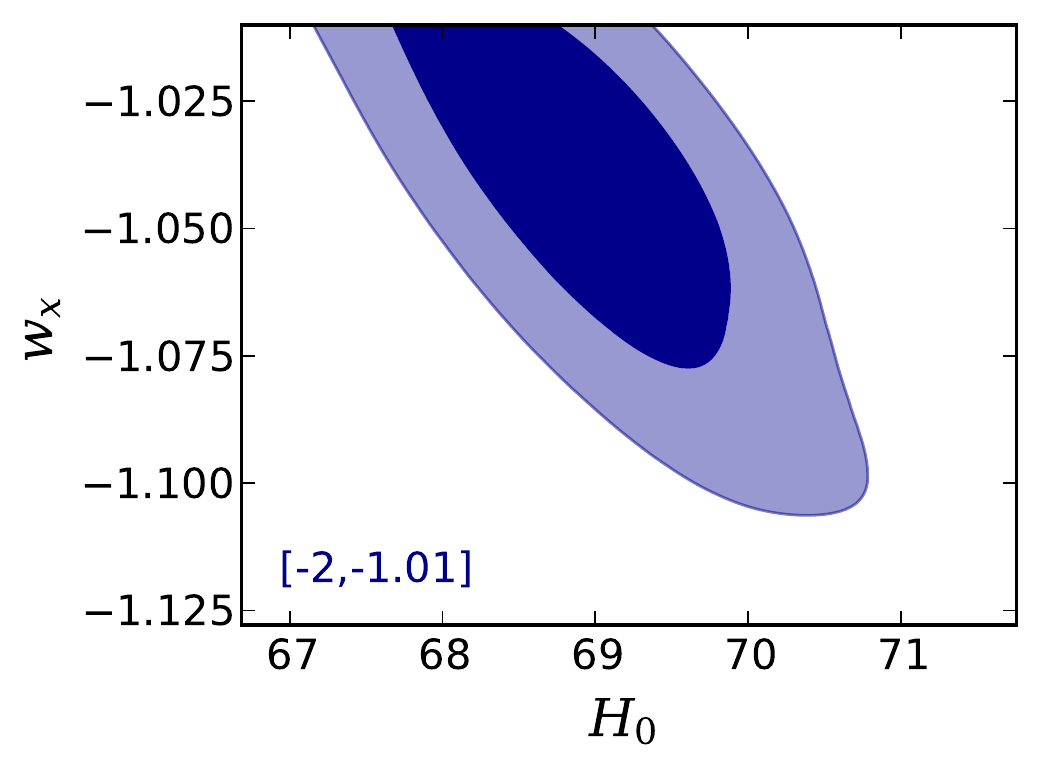}
		\includegraphics[width=0.24\textwidth]{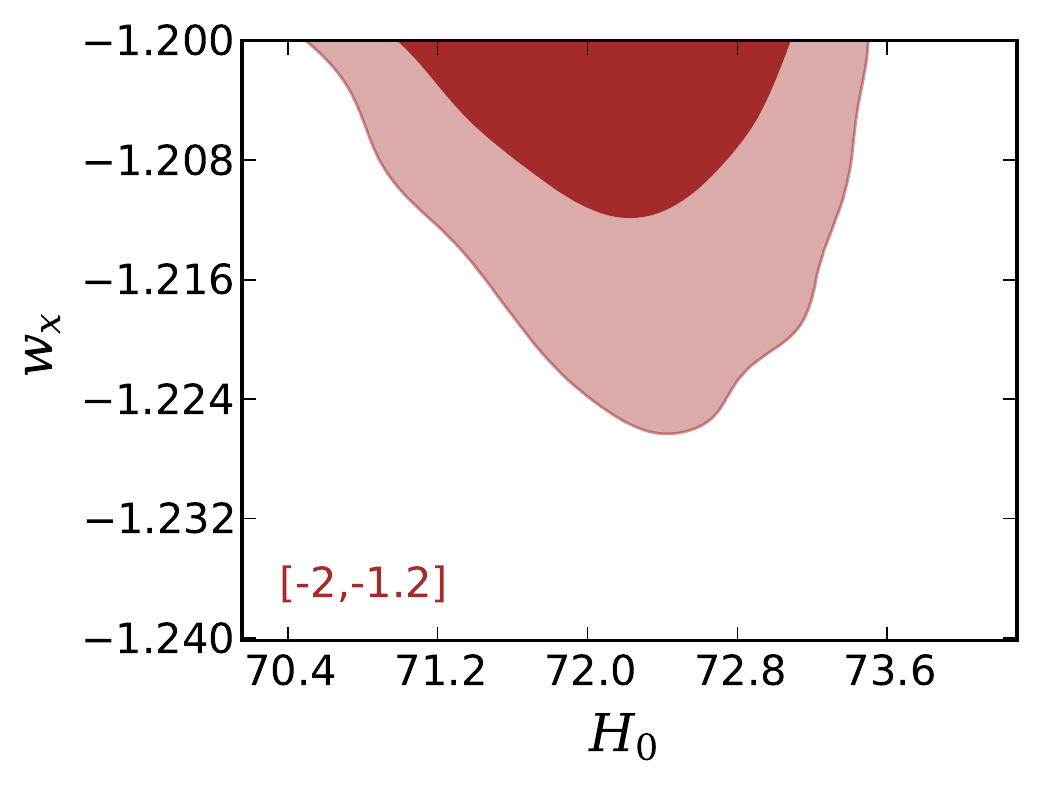}
		\caption{Contour plots for the parameters ($w_x$, $H_0$) considering four different regions of the constant dark energy equation of state. The analysis adopted here is the joint analysis from Planck TT, TE, EE $+$ lowTEB $+$ BAO $+$ JLA $+$ RSD $+$ WL $+$ CC $+$ R16. }
		\label{fig:tension1}
	\end{figure}
%\end{center}

%\begin{center}
\begin{table}%[tbh]
\caption{For the dynamical dark energy
equation of state, $w_x = w_0 +w_a (1-a)$, we present the results of the interaction model for four different regions of $w_0$, the current value of $w_x(a)$. The analysis has been performed for
the combined analysis Planck TT, TE, EE $+$ lowTEB $+$ BAO $+$ JLA $+$ RSD $+$ WL $+$ CC $+$ R16.
As mentioned earlier, $\Omega_{m0}= \Omega_{c0}+\Omega_{b0}$. }
\begin{tabular}{cccccccc}
\hline\hline
Parameters & $w_0 \in [-1.5, -0.9]$ &  $w_0 \in [-2, -1]$ & $w_0 \in [-2, -1.01]$ &$w_0 \in [-2, -1.2]$
\\ \hline
$\Omega_c h^2$ & $ 0.11670_{-    0.00120-    0.00583}^{+    0.00279+    0.00422}$ & $    0.11630_{- 0.00153- 0.00484}^{+    0.00266+    0.00423}$ & $0.11680_{-    0.00126-    0.00285}^{+    0.00159+    0.00289}$ & $0.11614_{-    0.00266-    0.00535}^{+ 0.00271+    0.00528}$ \\

$\Omega_b h^2$ & $ 0.02228_{-    0.00015-    0.00031}^{+    0.00015+    0.00030}$ & $ 0.02229_{-    0.00014-    0.00029}^{+    0.00015+    0.00029}$ & $0.02229_{-    0.00014-    0.00028}^{+    0.00015+    0.00029}$ &$    0.02225_{-    0.00017-    0.00035}^{+    0.00018+    0.00033}$\\

$100\theta_{MC}$ & $ 1.04069_{- 0.00036- 0.00064}^{+    0.00032+    0.00071}$ & $ 1.04072_{- 0.00033- 0.00061}^{+ 0.00032+ 0.00064}$ & $1.04070_{-    0.00030-    0.00060}^{+    0.00032+    0.00060}$ & $ 1.04071_{-    0.00038-    0.00072}^{+    0.00038+    0.00074}$\\

$\tau$ & $    0.06519_{-    0.01686-    0.03507}^{+    0.01664+    0.03306}$ & $    0.06610_{-    0.01802-    0.03025}^{+    0.01609+    0.03227}$ & $0.0666_{-    0.01650-    0.03087}^{+    0.01578+    0.03067}$ &$    0.06021_{-    0.01744-    0.03822}^{+    0.02037+    0.03488}$ \\

$n_s$ & $    0.97556_{-    0.00439-    0.00850}^{+    0.00448+    0.00869}$ & $    0.97578_{-    0.00428-    0.00825}^{+    0.00421+    0.00802}$ & $0.97613_{-    0.00414-    0.00737}^{+ 0.00399+    0.00777}$ & $ 0.97483_{-    0.00471-    0.01079}^{+    0.00577+    0.00938}$\\

${\rm{ln}}(10^{10} A_s)$ & $    3.07082_{-    0.03246-    0.06887}^{+    0.03230+    0.06407}$ & $    3.07179_{-    0.03482-    0.05927}^{+    0.03156+    0.06314}$ & $3.07275_{-    0.03112-    0.05991}^{+    0.03083+    0.05896}$ &$    3.06041_{-    0.03212-    0.07292}^{+    0.03982+    0.06591}$\\

$w_0$ & $ -1.02042_{-    0.05121-    0.14529}^{+    0.10065+    0.12042}$ & $   -1.07026_{-    0.01821-    0.10996}^{+    0.07026+    0.07026}$ & $-1.05620_{-    0.01190-    0.05875}^{+    0.04620+    0.04620}$ &$   -1.22159_{-    0.00307-    0.04641}^{+    0.02159+    0.02159}$\\

$w_a$ & $   -0.04104_{-    0.30556-    0.46861}^{+    0.22222+    0.54574}$ & $    0.10561_{-    0.20733-    0.34558}^{+    0.18514+    0.38729}$ & $0.07982_{-    0.11395-    0.25762}^{+    0.12230+    0.24061}$ &$    0.47808_{-    0.10169-    0.38255}^{+    0.19475+    0.29336}$\\

$\xi$ & $   -0.27803_{-    0.07809-    0.50761}^{+    0.27803 +  0.27803}$ & $   -0.19671_{-    0.03874-    0.33102}^{+    0.19671+    0.19671}$  & $-0.13835_{-    0.05235-    0.16374}^{+    0.13256+    0.13835}$ &$   -0.09762_{-    0.02265-    0.15496}^{+    0.09762+    0.09762}$\\

$\Omega_{m0}$ & $    0.29657_{-    0.00837-    0.02268}^{+    0.01197+    0.02164}$ & $    0.29318_{-    0.00926-    0.01932}^{+    0.01031 +    0.01949}$ & $0.29538_{-    0.00807-    0.01591}^{+    0.00877+    0.01602}$ &$    0.28267_{-    0.00818-    0.01598}^{+    0.00880+    0.01552}$\\

$\sigma_8$ & $    0.81234_{-    0.01528-    0.03984}^{+    0.02206+    0.04015}$ & $    0.81058_{-    0.01586-    0.03690}^{+    0.02005+    0.03590}$ & $0.81333_{-    0.01234-    0.02853}^{+    0.01479+    0.02705}$ &$    0.81078_{-    0.02158-    0.04294}^{+    0.02225+    0.04082}$\\

$H_0$ & $   68.63729_{-    0.88622-    1.46090}^{+    0.73591+    1.61172}$ & $   68.93131_{-    0.75984-    1.54095}^{+    0.77984+    1.47811}$ & $68.79597_{-    0.68156-    1.41596}^{+    0.71402+    1.48429}$ &$   70.14999_{-    1.06080-    1.70609}^{+    0.73942+    1.95652}$\\
\hline
\end{tabular}%
\label{tab:tension2}
\end{table}
%\end{center}

%\begin{center}
	\begin{figure}[tbh]
		\includegraphics[width=0.24\textwidth]{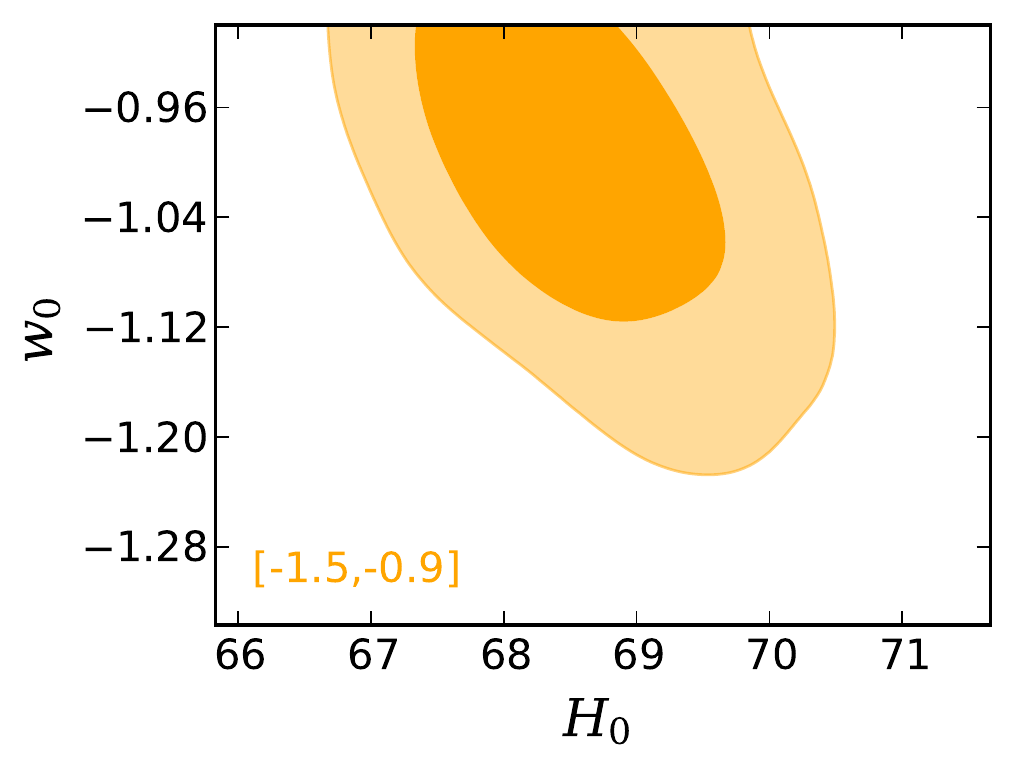}
		\includegraphics[width=0.24\textwidth]{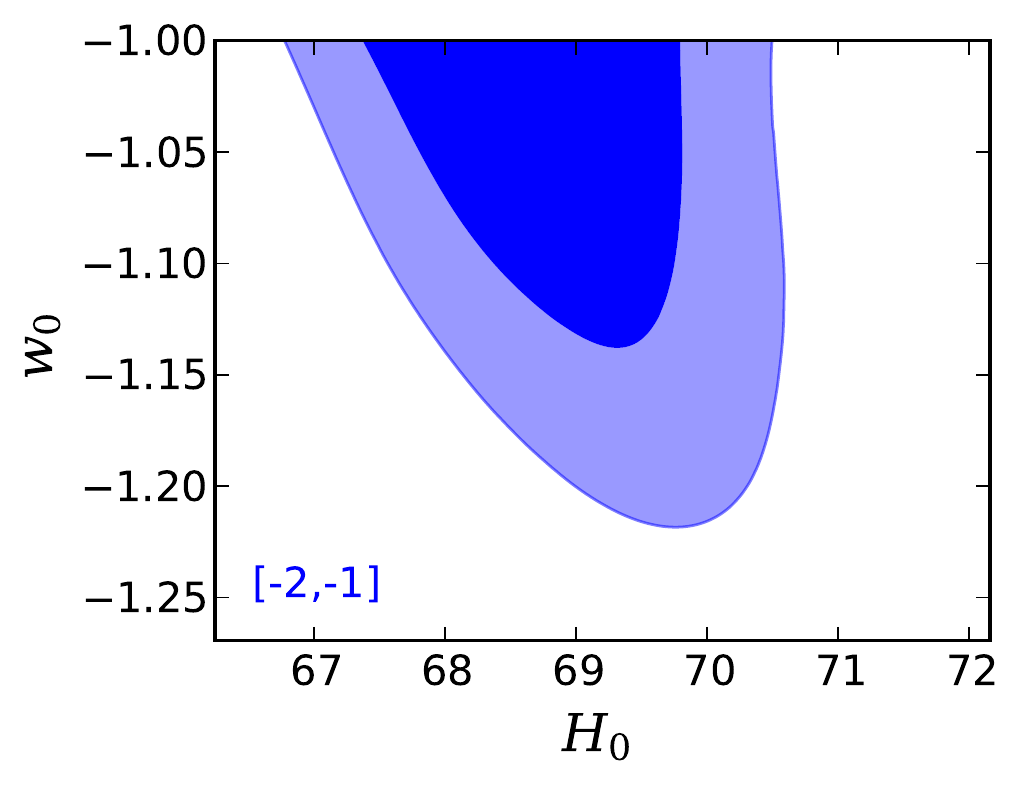}
		\includegraphics[width=0.24\textwidth]{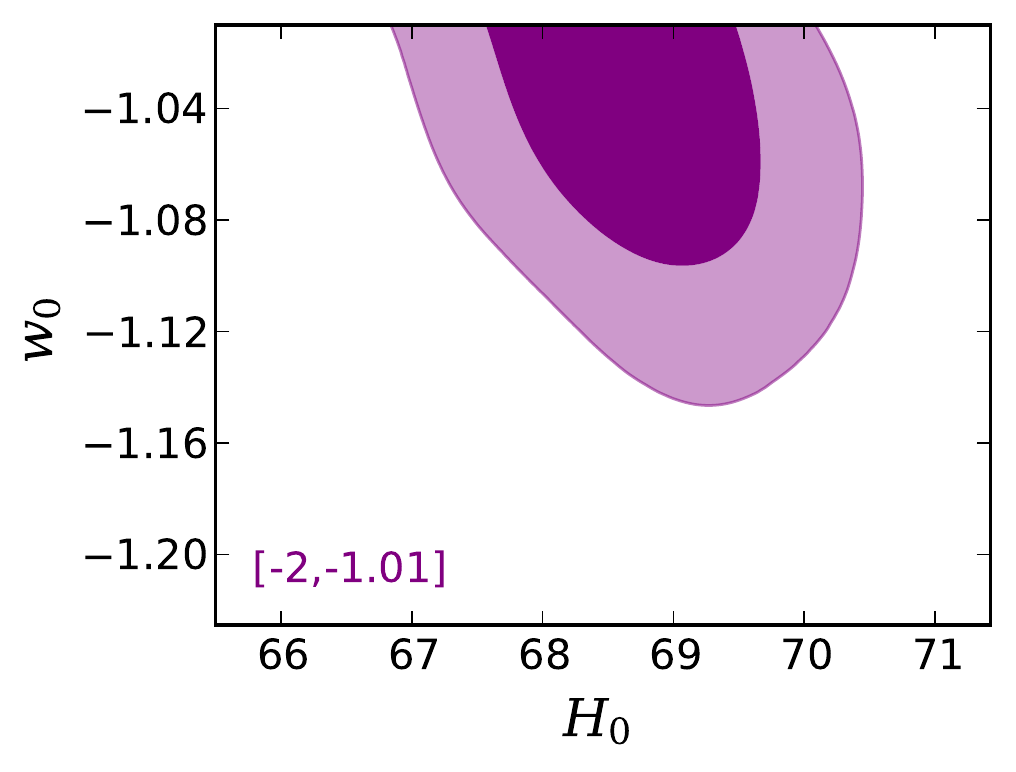}
		\includegraphics[width=0.247\textwidth]{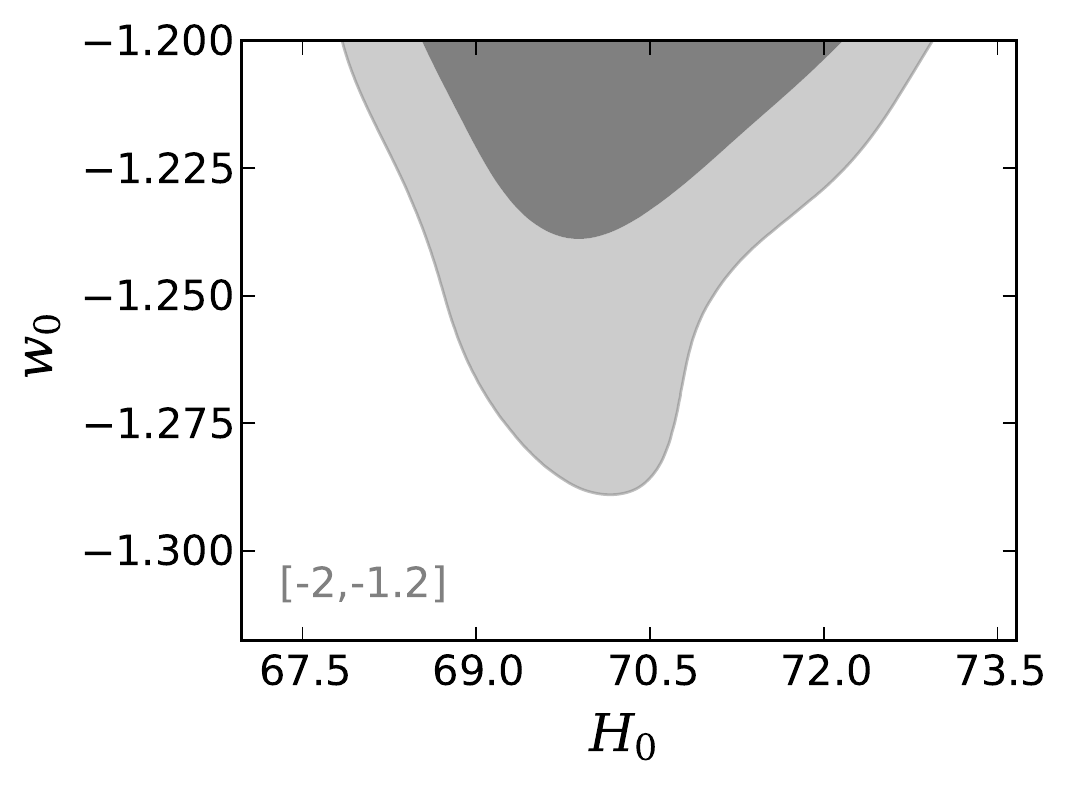}
		\caption{Contour plots for the parameters ($w_0$, $H_0$) considering four different regions of $w_0$, current value of the dynamical dark energy equation of state. The analysis adopted here is the joint analysis from Planck TT, TE, EE $+$ lowTEB $+$ BAO $+$ JLA $+$ RSD $+$ WL $+$ CC $+$ R16.}
		\label{fig:tension2}
	\end{figure}
%\end{center}

\end{itemize}

\section{Summary and Conclusions}
\label{conclu}

The stability analysis of interacting dark energy models generally divides the parameters space into two separate regions: (i) $w_x \geq -1$ and $\xi \geq 0$ or (ii) $w_x \leq -1$ and $\xi \leq 0$ in which $w_x$ is the state parameter for dark energy and $\xi$ is the coupling parameter of the interaction. That means, a discontinuity in the parameters space! A very recent study \cite{ypb} shows  that such difficulties concerning the discontinty in the parameters space can be solved with a new interaction that includes $(1+w_x)$ in the interaction function $Q$. Such construction of interacion function is motivated from the pressure perturbations for dark energy which informs that the inclusion of such factor releases the prior on the dark energy equation of state. That means, the factor ``$(1+w_x)$'' seems to play a vital role in the analysis of the interacting dark energy models at large scale of the universe. Motivated by this fact, in this work we introduce an interaction,
$Q = \xi \dot{\rho}_x$, which automatically contains a term $(1+w_x)$ [see eqn. (\ref{Q-1})] and consequnelty, this model holds the same features as in the  models of Ref. \cite{ypb}.

We constrained such typical interaction using the combined astronomical data  Planck TT, TE, EE $+$ lowTEB $+$ BAO $+$ JLA $+$ RSD $+$ WL $+$ CC $+$ R16. From Table \ref{tab:constantw} and Table \ref{tab:dynamicalw}, we see that both IDE 1 and IDE 2 allow a very small but nonzero interaction in the dark sectors, however, the zero value of the coupling parameter $\xi$ is not rejected at all. In fact, within $1\sigma$ confidence-level, $\xi = 0$, is consistent with the observational data. Further, the analysis also shows that the current value of the dark energy equation of state (both mean and the best fit) in both interacting models favors its phantom behaviour while in the $1\sigma$ confidence level, $w_x = -1$, is compatible with the observational data we employ. That means both
the interacting models exhibit a very slight deviation from
the $\Lambda$ cosmology while within $1\sigma$ confidence-region $\Lambda$
cosmology is consistent with the observational data. It is also interesting to note that at the perturbative level, the models are found to be very close to each other and also with the $\Lambda$-cosmology.

Our analysis also shows that for both IDE 1 and IDE 2, if the Hubble parameter values decrease, the dark energy equation of state, shifts its behaviour from phantom to quintessence. This behaviour was exactly observed in a recent work \cite{ypb}.  It is also interesting to remark that
the current IDE models can alleviate the tension
on $H_0$ that is observed in its global \cite{Ade:2015xua} and local
measurements \cite{Riess:2016jrr}. In fact, the theory of interacting dark energy
can be a reasonable direction to talk about the current tension on
$H_0$. We paid a considerable attention
on this issue with different regions of the dark energy equation of state
which concludes that the tension
on $H_0$ is reconciled in the phantom region while
the analysis also makes a note that the dark energy equation of
state with strong phantom nature needs further attention
to reach a definite conclusion!

\section*{ACKNOWLEDGMENTS}
The authors thank the referee for some important comments. W. Yang's work is supported by the National Natural Science Foundation of China under Grants No. 11705079 and No. 11647153. The work of SP was supported by the SERB-NPDF programme (File No: PDF/2015/000640). DFM acknowledges the support from the Research Council of Norway.

\end{document}